\documentclass[journal]{IEEEtran}
\ifCLASSINFOpdf
\else
\fi
\usepackage[utf8]{inputenc}
\usepackage{graphicx}
\usepackage{hyperref}
\usepackage[section]{placeins}
\usepackage[section]{placeins}
\usepackage{lscape}
\usepackage{amsmath,amsfonts,amssymb}
\usepackage{cuted}
\usepackage{xcolor}
\usepackage{caption} 
\usepackage{lscape}
\begin{document}
%
\title{Enhancing synthetic training data for quantitative photoacoustic tomography with generative deep learning}
%
%
%

\author{Ciaran~Bench$^{i}$, Ben~Cox$^{ii}$, \\ $^{i}$Ear Institute, University College London, London, UK \\$^{ii}$Department of Medical Physics and Biomedical Engineering, University College London, London, UK 
}

\maketitle

\begin{abstract}
Multiwavelength photoacoustic images encode information about a tissue's optical absorption distribution. This can be used to estimate its blood oxygen saturation distribution (sO$_2$), an important physiological indicator of tissue health and pathology. However the wavelength dependence of the light fluence distribution complicates the recovery of accurate estimates, in particular, preventing the use of a straightforward spectroscopic inversion. Deep learning approaches have been shown effective at producing accurate estimates of sO$_2$ from simulated data. Though, the translation of generic supervised learning approaches to real tissues is prevented by the lack of real `paired' training data (multiwavelength PA images of \textit{in vivo} tissues with their corresponding sO$_2$ distributions). Here, we discuss i) why networks trained on images simulated using conventional means are unlikely to generalise their performance on real tissues, and ii) the prospects of using two generative adversarial network based strategies to improve the generalisability of sO$_2$-estimating networks trained on synthetic data: a) CycleGAN-driven unsupervised domain adaptation of conventionally simulated images, and b) the generation of paired training data using AmbientGANs.   
\end{abstract}


%
\IEEEpeerreviewmaketitle

\section{Introduction}
\label{sect:intro}  

Information about a tissue's blood oxygen saturation (sO$_2$) distribution can be used to assess patient health and monitor tumour therapies. Therefore, there is a demand for a modality that can provide high resolution images of this haematological parameter \cite{cox2012quantitative}. Diffuse optical tomography, and Blood Oxygen Level Dependent Magnetic Resonance Imaging (BOLD MRI) can provide information about or related to sO$_2$. However the former only provides low resolution images at superficial depths while the latter is only sensitive to changes in blood volume and venous deoxyhaemoglobin concentration \cite{villringer1993near,ogawa1990brain}. 

In contrast, photoacoustic (PA) tomography can provide both superior resolution and sensitivity to both oxyhaemoglobin and deoxyhaemoglobin. Images are acquired by initially sending pulses of near infrared range (NIR) laser light into the tissue, where photons undergo several scattering events before being absorbed by chromophores. The subsequent relaxation of the excited chromophores raises the temperature of their surrounding environments inducing local increases in pressure that propagate to the sample surface as acoustic waves. Here, transducers (e.g. based on piezoelectrics or Fabry-Perot interferometers) record the waves as pressure time series. Recorded pressure time series can be used to reconstruct images of the photoacoustic initial pressure distribution ($p_0(x,\lambda)$ where $x$ is the location within the tissue, and $\lambda$ is the illumination wavelength) using one of several algorithms \cite{poudel2019survey,hauptmann2020deep}. Unlike purely optical modalities, information about the optical absorption is encoded in acoustic waves that propagate to the sample surface. Compared to photons, these undergo comparatively little scattering, resulting in an improvement in penetration depth. The amplitude of a perfectly reconstructed PA image is given by:
\begin{equation}
    p_0(x,\lambda) = \mu_a(x,\lambda)\Gamma(x)\Phi(x,\lambda),
    \label{eq:p0}
\end{equation}
where $\mu_a$ is the optical absorption coefficient, $x$ is the location within the sample, $\lambda$ is the optical illumination wavelength, $\Gamma$ is the PA efficiency, and $\Phi$ is the light fluence. 

It is clear from Eq. \ref{eq:p0} that PA images encode information about the sample's optical absorption coefficient. Knowledge of this parameter at multiple wavelengths (or $\mu_a$ scaled by a wavelength independent constant) can be used to quantify chromophore concentrations with a straightforward spectroscopic inversion (assuming all chromophore species and their molar absorption spectra are known) \cite{hochuli2019estimating}. However, acquiring accurate estimates of sO$_2$ from multiwavelength PA images is non-trivial for several reasons \cite{cox2012quantitative}. Chief among these is the confounding effect of the wavelength-dependent fluence distribution. A PA image can be viewed as an image of the $\mu_a$ distribution scaled by an (assumed) wavelength independent $\Gamma$ and a wavelength dependent $\Phi$. It is this fluence term that prevents the effective use of a simple spectroscopic inversion of multiwavelength PA image amplitudes to estimate chromophore concentrations. Spectral colouring is a term used to describe how the fluence alters a tissue region's PA spectra so that it is not directly proportional to the region's optical absorption spectra. 

If available, an accurate estimate of the multiwavelength fluence distribution can be used to correct the fluence term out of each image and restore the ability to recover concentration estimates with a spectroscopic inversion. In principle this could be acquired by running a simulation of light propagation in a synthetic model of the tissue. However, because tissue is highly scattering at NIR wavelengths, the spatial distribution of absorbers and scatterers throughout the tissue must be known to acquire an accurate estimate of the fluence. Therefore, the key challenge with quantifying chromophore concentrations from PA images is finding some way to compensate for the multiwavelength fluence distribution using only incomplete prior knowledge of the sample's optical properties. 

Despite the high errors that may occur from spectral colouring, sO$_2$ estimates acquired with an approximate linear inversion strategy \textit{without} an accompanying fluence correction (referred to here as sO$_{2}^{*}$) have been used to assess changes in a tissue's oxygenation or pathology state. In some of these studies the error due to spectral colouring is acknowledged, and the effective utilisation of sO$_{2}^{*}$ values is demonstrated by exhibiting how changes in this parameter with the onset of some physiological challenge are consistent with the expected change in tissue state \cite{kirchner2019photoacoustic, kirchner2019photoacoustics}. Though this approach should be used with caution as there is no guarantee that changes in sO$_{2}^{*}$ will reflect changes in the actual sO$_2$ or act as an equivalent biomarker in all cases. The degree of spectral colouring can change considerably with changes in tissue state caused by differences in blood flow or other processes that affect the concentrations of different chromophores.

Several strategies have been proposed to overcome the challenges with estimating the wavelength dependence of the fluence. Here, we discuss those that have been applied to physical phantoms or real tissues (and therefore have been tested in realistic imaging scenarios), and are noninvasive (an ultimately preferable option given the additional expertise required to use contrast agents or invasive light sources \textit{in vivo} and to minimise patient trauma).

In some cases, fluence estimates are acquired by constructing synthetic tissue models using assumed optical properties of the real tissue (e.g. based on literature values) or those measured at low resolution with an adjacent imaging modality (e.g. Diffuse Optical Tomography (DOT) or oblique-incidence diffuse reflectance measurements \cite{ranasinghesagara2010combined, bauer2011quantitative,li2011integrated}). However, given the natural variation in \textit{in vivo} tissue properties, it is unlikely that the sample's tissue components will have the same optical properties as those reported in the literature for a specific sample. Furthermore, published values for the optical properties for various tissue types are often measured from excised samples which may differ from their \textit{in vivo} counterparts because of changes in density and blood content. There is also no guarantee that every unique tissue type present in the sample or their precise location will be known \textit{a priori}, and therefore there is likely to be a significant degree of mismatch between a synthetic tissue model and its corresponding real tissue. As for the adjacent modalities, fluence estimates derived from DOT measurements have a depth dependent resolution practically limited to around 2-3 mm and thus inadequately capture the finer-scale heterogeneity of real tissue \cite{bauer2011quantitative, bentz2018superresolution}. Additionally, the modality requires complex hardware making less appealing to utilise. Sample optical properties have also been estimated from PA images by i) making the implicit assumption that light propagation can be accurately modelled in 1D, ii) assuming the tissue can be modelled as a series of optically homogeneous layers (each represented as a plane) and then iii) fitting a 1D light model (e.g. Beer-Lambert law) to the measured data. However, 1D light propagation models are only valid under a narrow set of conditions that are unlikely to be met in a real tissue \cite{cox2012quantitative}. Furthermore, given their typical heterogeneity most tissues are not accurately modelled as a series of planar layers. 

A method to measure the fluence non-invasively using ultrasonically tagged light has been proposed \cite{zemp2007ultrasound,daoudi2012correcting,daoudi2012towards,hussain2016quantitative}). However, this also assumes a significant degree of homogeneity in tissue properties that is unlikely to be met in most \emph{in vivo} imaging scenarios. 

Statistical separation/decomposition approaches such as those based on Independent Component Analysis (ICA) have been applied to images of synthetic phantoms \cite{an2018estimating}. However, ICA relies on the assumption that chromophores are spatially distributed in a statistically independent manner, which is certainly not the case for Hb and HbO$_2$ \textit{in vivo}.

Iterative error minimisation techniques (specifically model-based inversion) are another class of approaches that have been tested in phantoms and \textit{in vivo} tissue \cite{cox2006quantitative,Jiang2007Quantitative,jetzfellner2009performance, yin2007tomographic,zemp2009photoacoustic,laufer2006absolute,laufer2006quantitative}. Here, multiwavelength images of the sample ($x$) are acquired, and an accurate model of the imaging system is formulated (referred to here by the operator $I$). A synthetic tissue model (referred to as $c$) is used as an input to $I$ that outputs synthetic PA images $\hat{x}$, so $I(c) \rightarrow \hat{x}$. The parameters of $c$ are then updated so as to decrease the error between $\hat{x}$ and $x$. This process is repeated iteratively until the error between $\hat{x}$ and $x$ satisfies some criterion. The version of $c$ that satisfies this condition is assumed to accurately represent the properties of the real tissue/sample. This relies on $I$ being a highly accurate model of system used to acquire $x$. However, formulating an accurate model is challenging in practice due to the difficulty with precisely characterising all aspects of the signal acquisition pathway. This is one of several challenges with using this approach. Additionally, when the optical scattering distribution is unknown (and therefore recovered as part of the inversion) the inversion is ill-posed adding further complications. Also, all of the different chomophore species present in significant quantities in the sample (as well as their molar absorption/scattering spectra) must be known \textit{a priori}. This information may not always be available for \textit{in vivo} samples. Therefore, this strategy has not been widely implemented to estimate chromophore concentrations \textit{in vivo}.

In contrast, data-driven techniques based on deep neural networks do not require a well-characterised imaging system. Instead, a generic supervised learning approach requires a training set composed of multiwavelength PA images of each tissue and their corresponding ground truth distributions of chromophore concentrations. CNN-based networks learn to associate an arrangement of particular image features (representing tissue components or other aspects of the image such as reconstruction artefacts) to an sO$_2$ distribution. However, there is currently no straightforward way to acquire the ground truth sO$_2$ distribution \textit{in vivo}. 

Consequently, all reported uses of deep networks involve training datasets composed of simulated images or images of physical phantoms \cite{kirchner2018context, grohl2021learned, kirchner2021multiple, durairaj2020unsupervised, olefir2020deep, olefir2018bayesian, cai2018end, yang2019quantitative, luke2019net, jandhyala2021experimental, yang2019eda, chen2020deep, grohl2018confidence, nolke2021invertible}. However, networks trained on either type of data will not perform well on images of \textit{in vivo} tissue. This is because the frequency at which modelled tissue components are found in each tissue, the features that represent them, how they are arranged in space, and the relation these image features have to the sO$_2$ may differ. Specifically, simulated/phantom datasets belong to a data domain that differs from that describing images of \textit{in vivo} tissue (see Section \ref{sec:domain_gap}).

Unsupervised learning techniques that do not require paired data have been tested on simulated datasets, but have yet to be applied to physical phantoms or real tissues \cite{durairaj2020unsupervised}. Visual transformer-based architectures may be effective at extracting and utilsing global contextual features relevant to learning sO$_2$ quantification, and have been used in unsupervised learning frameworks to perform registration of biomedical image data \cite{chen2021vit}. However, it is not clear how the scheme could be adapted to perform unsupervised image-to-image regression tasks. Therefore, acquiring a realistic dataset remains a critical step towards translating the use of deep neural networks to images of real tissue (and more broadly to validate any other proposed approach \textit{in vivo}). It may be possible to measure the sO$_2$ in several discrete regions of a tissue using inserted electrodes or probes, and then interpolate between them to estimate a continuous distribution to use as a ground truth \cite{gehrung2019development,sonmezoglu2021monitoring}. However, this is time consuming, costly, and would ultimately provide only a low resolution ground truth. Synthetic images are comparatively more cost effective to generate, and so it would be preferable to instead find a way to leverage this kind of data.

\section{The domain gap between real and simulated images of tissue}
\label{sec:domain_gap}
A network's ability to generalise its performance on a test set of data separate from its training set depends on the similarity of the two datasets' data domains. A data domain $D = \{\chi, \Upsilon, P(x,y)\}$ consists of an input feature space $\chi$ (a vector space containing all image features), an output feature space $\Upsilon$, and a joint probability distribution $P(x,y)$ over the input and output feature space pair $\chi \times \Upsilon$, where $x$ is an instance of the network inputs $x_1, x_2, ... x_i \in \textbf{x}$ and $y$ is an instance of the corresponding ground truths $y_1, y_2, ...y_i \in \textbf{y}$ \cite{kouw2019review, yang2009heterogeneous, csurka2017domain, pan2009survey, zhao2020review, toldo2020unsupervised, kouw2018introduction,liu2022deep}. In a supervised learning framework, training data will consist of image pairs $\{x_i , y_i\}$. The joint probability distribution can be decomposed into marginal (commonly referred to as the `data distribution') and conditional distributions: $P(x,y) = P(x)P(y|x)$ or $P(x,y)=P(y)P(x|y)$.

A network's performance will generalise well to any unseen test data when the domain describing the training set ($D_{train}$) is approximately equal to the domain describing the test data ($D_{test}$): $D_{train}\approx D_{test}$. Therefore, a network trained on simulated images may generalise its performance to images of real tissues if their domains are sufficiently similar. However, images generated using conventional simulation pipelines do not belong to the same domain describing images of real tissues for several reasons. Synthetic tissue models used in simulations are often constructed by approximating tissue components with simple shapes arranged in a way that appears to mimic components found in images of real tissues (acquired with various modalities). Optical and acoustic properties may be assigned based on values reported in the literature. The image data is then simulated by applying a forward model of image generation that includes a model of light transport, acoustic propagation, and image reconstruction using the synthetic tissue as an input. Though in some cases, only a light model is used to construct the forward model. In principle, it is possible to optimise the forward model by carefully characterising a known imaging system. However, conventionally simulated images may still fail to achieve domain-alignment with real images as the construction of sufficiently realistic tissue models requires highly accurate prior knowledge of the way optical properties are typically distributed in real tissues - information that is not readily available. 

\subsection{Causes of the domain gap}
Here we outline the causes of the domain gap between simulated training images, and test images of real \textit{in vivo} tissues. 
\subsubsection{Unequal feature spaces: $\chi_{train} \neq \chi_{test}$ or $\Upsilon_{train} \neq \Upsilon_{test}$}
If the shapes of tissue components are not sufficiently realistic, nor all of the processes involved with simulating image acquisition, then simulated images may contain features not found in images of real tissues. Consequently, a network trained on these simulated images will not be able to accurately detect tissue components (or other features such as artefacts) in real images. Ultimately, the network will have inadequate information about image contents, and consequently, sO$_2$ estimates will have low accuracy. Network accuracy may also suffer if the output feature space of the test set is not equal to the output feature space of the training set. 

\subsubsection{Unequal data distributions: $P(x)_{train} \neq P(x)_{test}$}
The frequency at which certain features representing arrangements of tissue components or artefacts occur in simulated images may differ from real images. This can happen if the synthetic tissue models in the training set are not sufficiently representative of the tissues depicted in the test set, or if the model used to simulate them does not reflect the imaging system used to acquire the real test images. For example, a network may be trained on a simulated dataset where only a few example tissues have melanin. If this network was applied to a test set of images depicting real tissues with melanin, the network would be unlikely to generalise its performance. This is because it has only learned from a few examples containing melanin, and so will not have learned how its presence may affect the fluence in a wider range of scenarios. 

\subsubsection{Unequal joint distributions: $P(x,y)_{train} \neq P(x,y)_{test}$}

It is possible for the same feature to occur in both the training and test set, but have a different relation to the ground truth depending on which dataset it is found in. For example, wing-like reconstruction artefacts in real images may be mistaken as vessel-like structures by a network that was trained on PA images simulated without an acoustic model. 

\subsection{Minimising the domain gap with transfer learning}
Transfer learning may provide a suitable framework for improving a network's performance on out-of-domain test data. Transfer learning refers to a class of techniques that enable a network trained to perform some `source' task (using data belonging to a source domain) to perform a related `target' task on data described by a target domain. More specifically, consider a network $K$ that is trained in a supervised manner on image pairs $\{\hat{x}_{i}, \hat{y}_{i} \} \in$ $\cal D$$_s \{ \chi_s , \Upsilon_s , P(\hat{x},\hat{y}) \}$ (where $\cal{D}$$_s$, is the source domain, with network inputs given by $\hat{x}_1, \hat{x}_2, ... \hat{x}_i \in \hat{\textbf{x}}$,  and outputs $\hat{y}_1, \hat{y}_2, ... \hat{y}_i \in \hat{\textbf{y}}$), to predict $\hat{y}_i$ for a corresponding input $\hat{x}_i$. In other words, it learns a mapping  $K: \chi_s \rightarrow \Upsilon_s$, which is referred to as the source task $\cal{T}$$_s$ \cite{kouw2019review,weiss2016survey}. Transfer learning aims to use $K$ in some form to perform the separate but related target task $\cal{T}$$_t$ of predicting the outputs $\tilde{y}_i$ from inputs $\tilde{x}_i$ belonging to the target domain $\cal{D}$$_t\{ \chi_t , \Upsilon_t , P(\tilde{x},\tilde{y}) \}$ where $\cal{D}$$_t\neq \cal{D}$$_s$. 

We are interested in using a network trained to perform sO$_2$ quantification on simulated PA images of tissue to perform the same task on images of real tissues. The most generic transfer learning strategies involve some kind of fine-tuning procedure, where a network pretrained on data from the source domain is trained on a small dataset belonging to the target domain. However, we are principally interested in unsupervised transductive transfer learning (of the heterogeneous variety) given i) the lack of available ground truths in the target domain, ii) the source and target tasks are the same, and iii) feature spaces and the data distributions of each domain are unequal \cite{pan2009survey}. Out of several possible strategies within this subset of transfer learning, we pursue the use of adversarial domain adaptation strategies. This is because adversarial networks offer an effective framework for automatically detecting and adapting relevant features to achieve domain alignment \cite{toldo2020unsupervised, farahani2020concise}. This is advantageous, as it is often the case that we do not know the kinds of features that are relevant to adapt \textit{a priori}, nor a hand-engineered and consistent method to detect these features and perform the required adaptation.

\subsection{Outline of proposed GAN-based domain alignment strategies}
\begin{figure*}[htp!]
\begin{equation}
\min_{G}\max_{D}V(G,D) = \min_{G}\max_{D}\mathbb{E}_{x\sim P_r}[\log D(x)] + \mathbb{E}_{z\sim P_z}[\log(1-D(G(z)))].
\label{eq:GAN_Loss}
\end{equation}
\end{figure*}
\subsubsection{Introduction to Generative Adversarial Networks}
Generative Adversarial Networks (GANs) provide a framework for implicitly learning how to randomly sample from a dataset's data distribution. The generic architecture is composed of two modules: a generator $G$ and discriminator $D$. $G$ takes noise (sampled from a Gaussian or uniform distribution) as an input, and outputs an image that (ideally) is indistinguishable from images found in the network's training set. $D$ takes an image (either produced by $G$ or from the training set) and classifies whether the image is synthetically generated (i.e. produced by $G$) or not. $G$ and $D$ are trained as adversaries; $G$ attempts to produce images that fool $D$, while $D$ is trained to differentiate synthetic and real images as accurately as possible. 

The task is described by \cite{saxena2021generative,creswell2018generative} and the objective function is given by Equation \ref{eq:GAN_Loss}. The first term represents the discriminator's predictions on real data ($x$ sampled from the training set's data distribution $P_r$), and the second refers to its predictions on fake/generated data (generated by inputting noise $z$ sampled from a uniform or Gaussian distribution $P_z$).

Generic GANs are notoriously challenging/unstable to train. One reason for this is because the distance metric implicitly used to assess the quality of the learned data distribution (Jensen-Shannon Divergence) is not always suitable for data distributions describing natural datasets \cite{arjovsky2017wasserstein,gulrajani2017improved}. In practice, more stable GAN variants such as the Wasserstein-GAN with gradient penalty (WGAN-GP) can be used instead. Some synthetic PA images generated with a WGAN-GP are provided in Appendix \ref{sec:Appendix_WGAN}. 

In this article, we use two variants of the generic GAN: an AmbientGAN, and a CycleGAN. 

\subsubsection{Ambient Generative Adversarial Networks}
A generic GAN can generate images that appear to belong to its training set. GANs have been used to facilitate the generation of training data for networks trained to estimate chromophore concentrations. In \cite{schellenberg2021data}, a GAN was used to generate random tissue models that appeared to belong to a set of simulated tissues constructed using the shapes of tissue components depicted in real PA image data. These generated models were then assigned optical properties, and used as an input to a forward model to simulate each tissue's corresponding PA images.

However, because we don't have the ability to acquire detailed information about the optical/acoustic properties of any given real tissue to provide to a GAN as training data, we can not use this approach to generate paired datasets of PA images of tissues with highly accurate and realistic ground truth information about their optical properties (i.e. this approach is inherently limited by our prior knowledge about how optical properties are typically distributed in real tissues). So how may we generate realistic images of tissue as well as its underlying tissue model without providing the network any examples of what an underlying tissue model looks like? Here, we describe how ambient generative adversarial networks (AmbientGANs) may provide the framework for learning this task \cite{zhou2020learning,bora2018ambientgan}.

The architecture of an AmbientGAN is similar to a generic GAN - the key difference is that a frozen, differentiable forward model of image generation $O$ is appended to the generator $G$. Instead of generating the synthetic image directly, $G$ now generates an input (a synthetic tissue model) to $O$ that then outputs an image indistinguishable from those in the training set. The architecture can be trained in the same manner as a generic GAN. $O$ must be expressed in a way so that it can be computed efficiently (it will be run for every batch), and so it is differentiable (allowing $D$ to provide feedback to $G$ through $O$). Though, there is no guarantee that a synthetic tissue model generated by $G$ that produces a realistic image after being processed by $O$ will itself reflect the properties of a real tissue. E.g. esoteric arrangements of scatterers can guide light towards absorbers in a way that produces an image indistinguishable from images of real tissues. Therefore, additional constraints are needed to ensure the generated tissue models are sufficiently realistic. 

\subsubsection{Cycle consistent generative adversarial networks}
Cycle consistent generative adversarial networks (CycleGANs) are another class of GAN that can adapt an image that belongs to one domain so that it appears to belong to another while preserving much of the inherent structure in the original image \cite{zhu2017unpaired}. This can be performed without the use of image pairs, lending itself well to unsupervised domain adaptation tasks. This could be used, for example, to adapt images simulated using conventional approaches to make them appear more similar to images of real tissues (i.e. align their feature spaces) in an unsupervised manner.

The architecture is composed of two generators and two discriminators - one for each of the data domains $X$ and $Y$. One generator $G$ adapts an image $x \in X$ so that it appears to belong to domain $Y$ as determined by the discriminator $D_y$ that classifies whether an image belongs to $Y$. The other generator $F$ adapts an image $y \in Y$ so that it appears to belong to $X$ as determined by the discriminator $D_x$ that classifies whether an image belongs to $X$.

Each training iteration involves several steps and loss functions. A cycle loss (the error between an image $x$ and $F(G(x))$ and the error between an image $y$ and $G(F(y))$) is used to ensure much of the original structure is preserved in the adapted images. An identity loss (error between $x$ and $F(x)$ and between $y$ and $G(y)$) is used to preserve inter-channel features/structure (e.g. colour for generic image adaptation tasks, or the wavelength dependence of the image amplitude in our case) in the adapted images. Generic discriminator and generator losses are also employed for each module. 

While training, first, images $x$ and $y$ are adapted. Then, the parameters for calculating the cycle-loss and identity loss are computed (i.e. $F(G(x))$ and $G(F(y))$. Subsequently, the adapted and original images are fed into their respective discriminators (i.e. $D_x(x)$, $D_x(F(y))$, $D_y(y)$, and $D_y(G(x))$). All the losses are calculated, and the generators and discriminators updated. Further details are provided in Appendix \ref{sec:cycleGAN_train}.

CycleGANs have been used to improve the performance of networks trained to estimate chromophore concentrations from simulated PA images when applied to real data \cite{li2022deep}. A CycleGAN (dubbed SEED-Net) was trained to adapt simulated $p_0$ images (of synthetic tissues modelled after a variety of samples, such as mouse muscle tissue and brain tissue) so that they appeared to belong to the same domain as real images (agar phantoms with absorbing/scattering inclusions of basic shapes, and images of \textit{ex vivo} tissue). In essence, the CycleGAN was trained to help align the input feature spaces of the experimental and simulated domains. A secondary dual-path network (QOAT-Net) based on the U-Net was then trained on adapted simulated images to estimate each sample's $\mu_a$ distribution (not adapted in any way) as ground truths. The trained network was then tested on real agar phantom images, and on real images of \textit{ex vivo} tissue phantoms also not used for training (though, only a limited set of examples are presented in the paper). Estimates of $\mu_a$ acquired from the real test images were more accurate compared to those acquired using QOAT-Net trained on non-adapted simulated images alone. When provided images of phantoms constructed from \textit{ex vivo} porcine liver tissue and tenderloin, the estimated $\mu_a$ for each tissue type were within the expected range, though, these results were not precisely validated. Their QOAT-Net was also applied to an \textit{in vivo} image of a mouse cross section, where the estimated $\mu_a$ was within the expected range. 

Although this work shows how adapting network inputs with adversarial training strategies can improve the accuracy of $\mu_a$ estimates acquired from experimentally acquired target data, the successful application of this approach on real tissue remains uncertain for a few reasons. Firstly, SEED-Net only learns to make the input feature spaces of the source and target domains more similar. Consequently, QOAT-Net only learns to output oversimplified $\mu_a$ distributions. Despite receiving a realistic looking input, it may not be able to output $\mu_a$ distributions with properties typical of those found in the real phantoms. Furthermore, the only constraint on the adaptation performed by SEED-Net is that the input images appear indistinguishable from the real set of images. QOAT-Net would ideally learn to invert the model of image generation associated with the target (real/phantom) images. Therefore, an ideal adaptation would be constrained to ensure that adapted images are those that would have been produced from the same underlying phantom/tissue if the model of image generation for target images were to be applied to it. Without this constraint, there is no guarantee that QOAT-Net will be learning to invert the forward model underlying the acquisition of the target images. However, it is not clear as to how such a constraint could be formulated.

\section{Outline of experiments}
Here, we discuss the potential of two GAN-based strategies for improving the quality of simulated training data: 
\begin{enumerate}
    \item First, we describe a toy example showing how CycleGANs can be used to improve the generalisability of an sO$_2$-estimating network by aligning the input feature spaces of a source domain of images simulated with a light propagation model, an acoustic model, and an image reconstruction model and a target domain composed of images simulated with a light model alone. 
    \item We also provide a demonstration of how AmbientGANs can be used to generate paired training data for chromophore quantification using only PA images as inputs. We generate PA images of circular absorbers in a homogeneous absorbing background as well as images of their underlying optical absorption distributions. 
\end{enumerate}

\section{Methods and Results}
\subsection{Unsupervised domain adaptation with a CycleGAN}
\label{sec:method_cyclegan_uda}
\begin{figure*}
    \centering
    \includegraphics[width = 1.4\columnwidth]{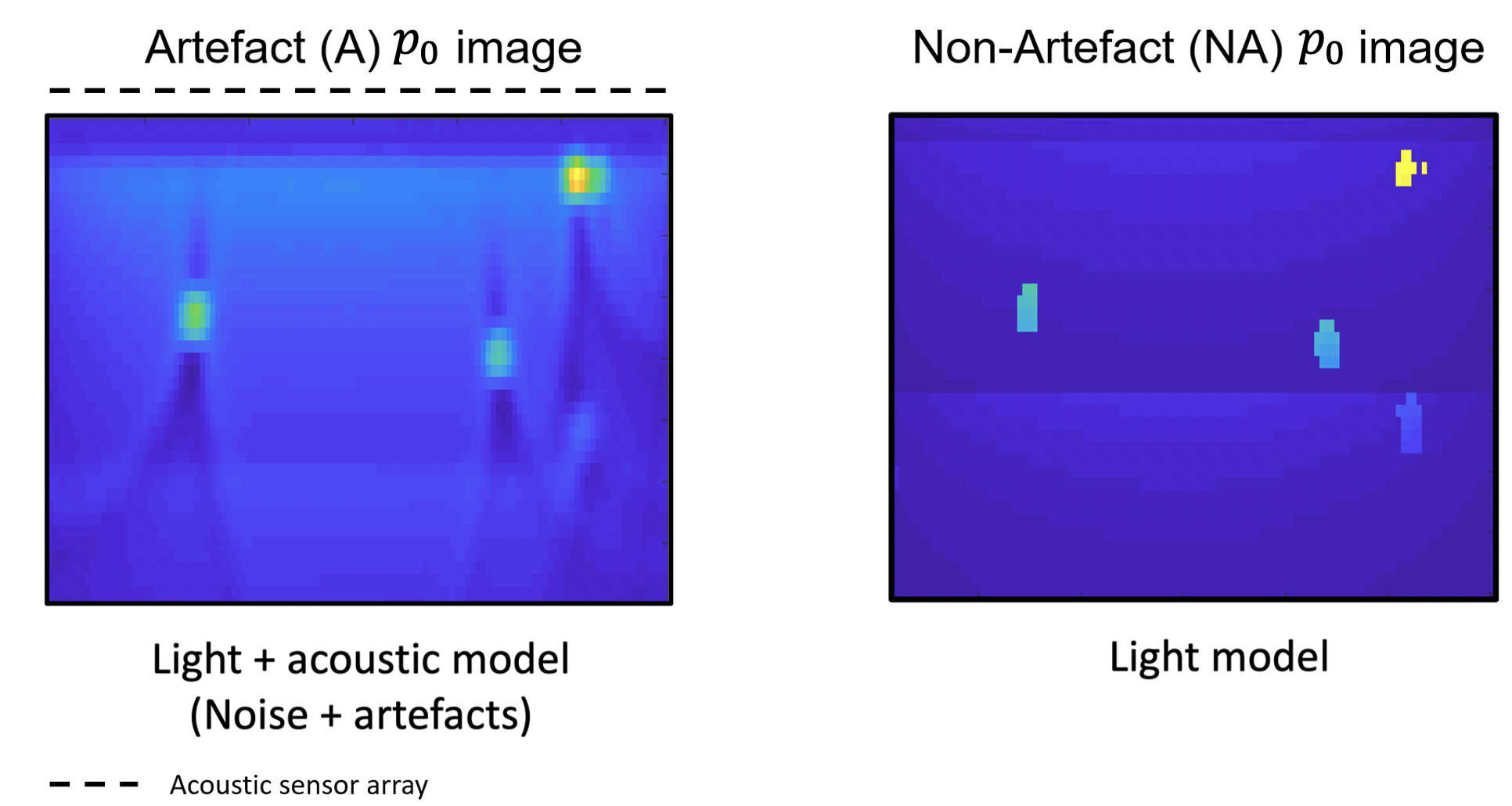}
    \caption{An example of a 2D slice of an image with reconstruction artefacts ($A$) produced using a simulation pipeline consisting of a light, acoustic, and image reconstruction model along with the same tissue's corresponding non-reconstructed image without artefacts (\emph{NA}) simulated with a light model alone. Images from these two domains were used in the CycleGAN-based domain adaptation experiment. This figure is used to demonstrate the differences in the images from each domain as opposed to showing training examples fed to the CycleGAN.}
    \label{fig:R_NR_examples}
\end{figure*}

\begin{figure*}
    \centering
    \includegraphics[width = 1.7\columnwidth]{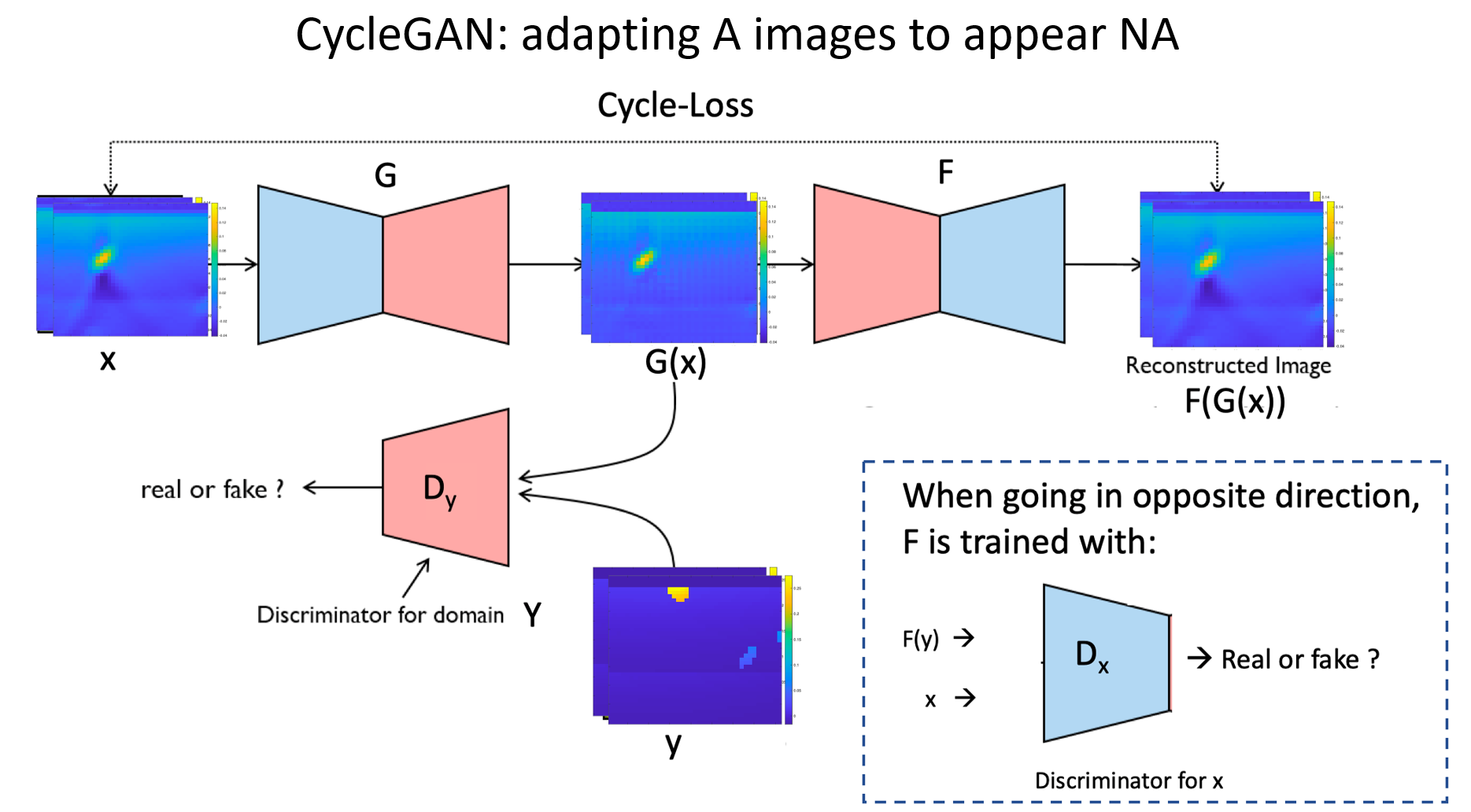}
    
    \caption{Schematic showing how data flows through the CycleGAN in the case where reconstructed ($A$) images ($x$ belonging to domain $X$) will be adapted to appear non-reconstructed (\emph{NA}) ($y$ belonging to domain $Y$). The generator $G$ adapts $A$ images from domain $X$ to appear as though they belong to domain $Y$ (i.e. appear like an \emph{NA} image).  Generator $F$ adapts an image belonging to domain $Y$ so that it appears to belong to domain $X$. The Cycle loss (difference between the input $A$ image $x$, and its reconstructed counterpart $F(G(x))$ ) is used to ensure the adaptation preserves much of the original structure found in $x$. Discriminator $D_y$: determines whether an input is a real or generated example from the $Y$ domain. Discriminator $D_x$: determines whether an input is a real or generated example from the $X$ domain.}
   
    \label{fig:cycleGAN_structure}
\end{figure*}

\begin{figure*}
    \centering
    \includegraphics[width = 1.99\columnwidth]{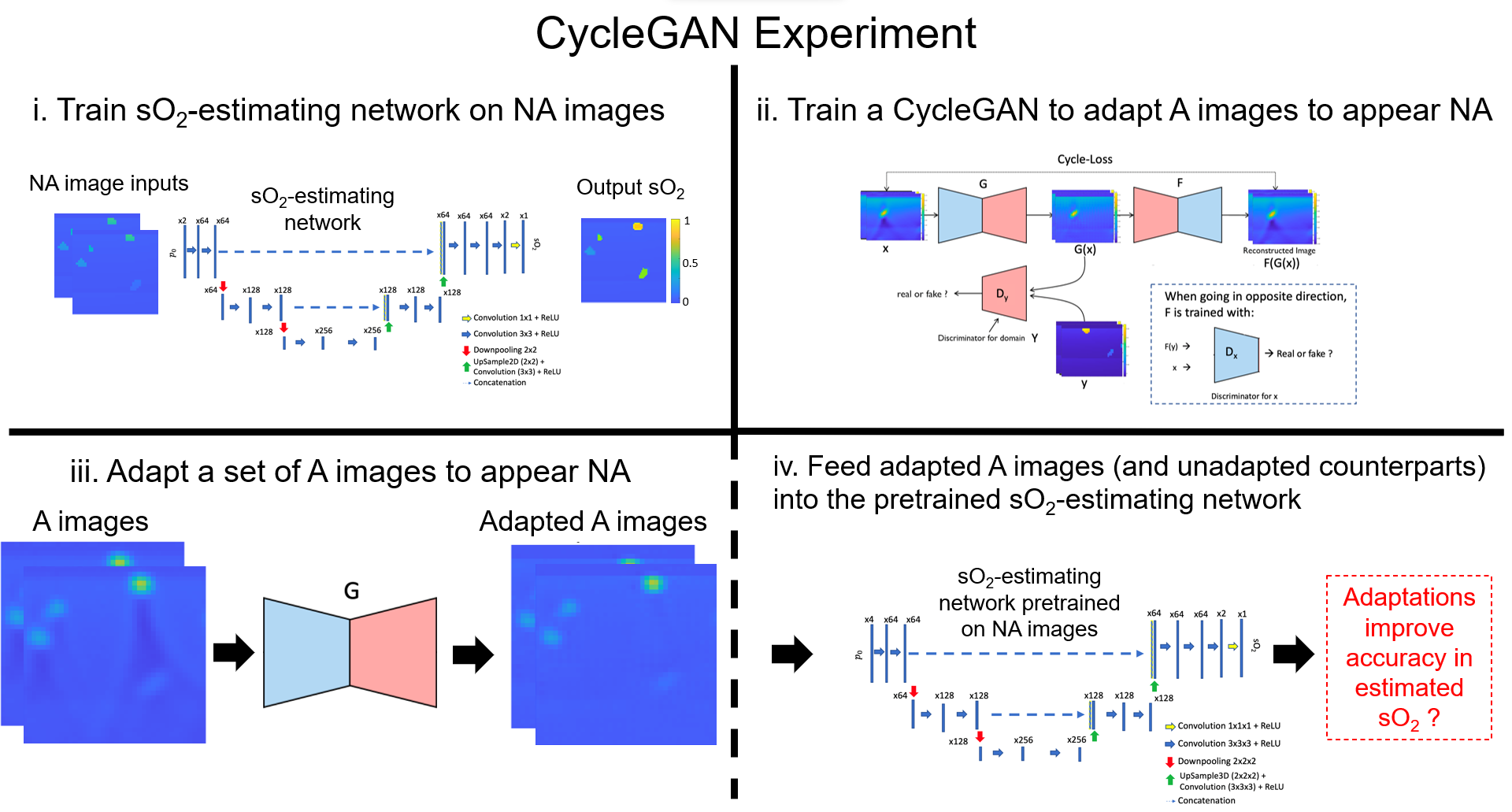}
    \caption{Summary of the CycleGAN domain adaptation experiment. Adapted $A$ images were used as inputs to an sO$_2$-estimating network pretrained on \emph{NA} images, and the resulting outputs were compared to those produced from using the corresponding unadapted images as inputs.}
    \label{fig:cycleGAN_summary}
\end{figure*}

We sought to investigate how CycleGANs could be used to improve the performance of an sO$_2$-estimating network on out-of-domain target data with domain adaptation. An ideal test would involve the use of simulated source data with real target data (e.g. multiwavelength PA images of real tissues with their corresponding ground truth sO$_2$ distributions). However, given the challenges with acquiring information about the ground truth sO$_2$ distributions in real tissues, we instead opted to conduct a study utilising simulated data. This allowed us to assess whether this approach may be effective even in the highly idealised case where any domain mismatch arises from noise and reconstruction artefacts alone. A summary of the experiment is shown in Fig. \ref{fig:cycleGAN_summary}.

We consider two datasets: i) a dataset composed of 2D PA images of vessels immersed in multi-layer tissues simulated using a forward model that considered light propagation, acoustic propagation, and image reconstruction with added noise following the procedure described in \cite{bench2020toward} (referred to here as artefact `$A$' images) and ii) another set of images simulated with a light model alone (referred to as non-artefact or `\emph{NA}' images). An example of each image type is shown in Fig. \ref{fig:R_NR_examples} (though, this figure does not depict a training example). To demonstrate whether this approach may work even in the ideal case where there is partial alignment in the source and target domains, each dataset was simulated to ensure that any domain mismatch would arise from the presence of artefacts and noise in the $A$ images.

\emph{NA} images were simulated by i) running a light model (as described in \cite{bench2020toward}) at two illumination wavelengths (784 nm and 820 nm) for all examples in a set of 3D tissue models (dimensions $40\times120\times120$ pixels, corresponding to real dimensions of $4 \times12 \times12$ mm), then ii) multiplying each fluence distribution by their respective tissue model's corresponding $\mu_a$ distribution, then iii) parsing a 2D slice (slice 60 from the second dimension, resulting in a $40 \times 120$ pixel image) from the resulting image volumes, and lastly iv) cropping the first and last 40 columns from each image, resulting in dimensions of $40 \times 40$ pixels. $A$ images were simulated by performing steps i) - ii), and then following this with the application of acoustic model and image reconstruction model (as described in \cite{bench2020toward}) to each image volume. A 2D slice (slice 70) was then parsed from each reconstructed image volume and cropped in the same manner as the \emph{NA} images to share the same dimensions. Parsing $A$ images from slice 70 of the 3D image volumes ensured that each example from each domain depicted different contents. Slices near the middle of each tissue model were chosen to ensure images were likely to contain a dense arrangement of vessels.

The domains describing either set of data were similar in the following ways. Firstly, given the image slices were taken from central regions of each tissue, the ground truths are composed of dense arrangements of vessels and consequently, the output features spaces will have a significant degree of alignment. The joint probability distributions and input feature spaces may also share some degree of alignment as the illumination parameters of the light model were identical for each image set and images were parsed from different (but central) regions of the same underlying 3D tissue models. Despite these similarities, there was considerable mismatch in the input feature spaces and joint probability distributions due to noise and reconstruction artefacts. The potential consequences of this are shown by the poor generalisation achieved by an sO$_2$-estimating network trained on $A$ images when applied to \emph{NA} images as shown in Appendix \ref{sec:Appendix_poor_generalisation}. 

A CycleGAN (see Fig. \ref{fig:cycleGAN_structure} for a schematic and Fig. \ref{fig:cycleGAN_arch} for architecture details) was trained on 400 2D images from each set for 6 epochs. The stopping point was determined by visually inspecting a validation set of 10 $A$ images and terminating training when most of the wing-like reconstruction artefacts, along with the low-amplitude artefact region commonly found just beneath vessels appeared to have been removed from the images (further training details are given in Appendix \ref{sec:cycleGAN_train}). The earliest possible epoch satisfying our heuristic criteria was chosen to prevent the images from being `overadapted' and therefore less representative of their underlying tissue. A more quantitative quality metric would be preferable to improve the consistency and the robustness of our stopping criteria. Furthermore, a larger validation set would have been preferable. However, due to constraints on the amount of available training data, the use of a larger validation set was not possible. 


\begin{figure*}[]
    \centering
    \includegraphics[width = 1.7\columnwidth]{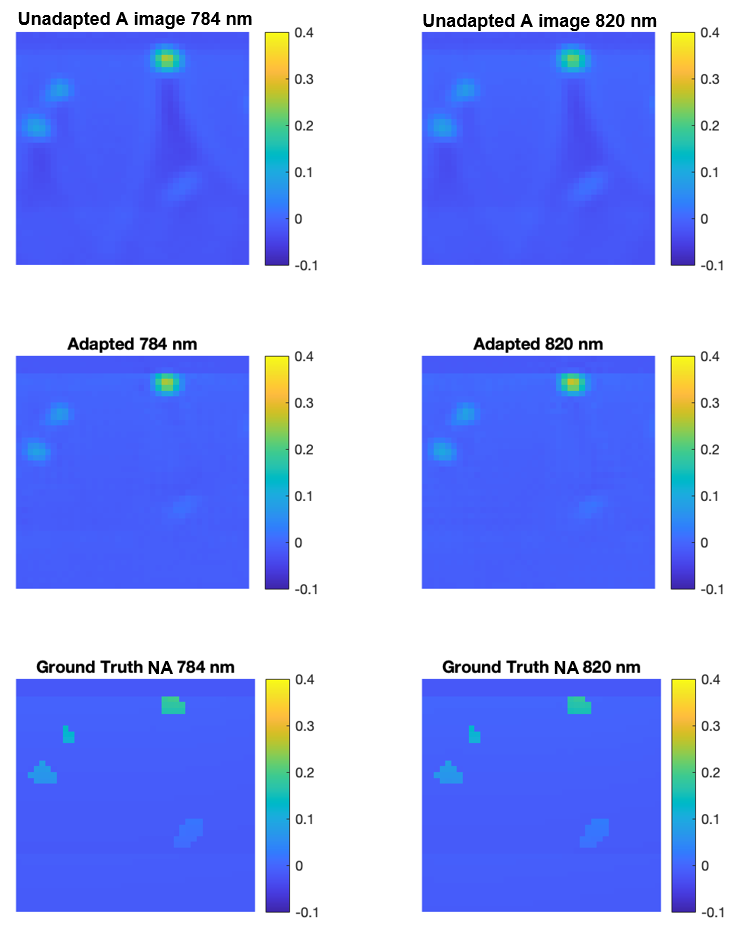}
    \caption{Example of two unadapted $A$ images (top row), their adapted counterparts appearing as though they belong to the \emph{NA} domain (middle row), and the corresponding `true' \emph{NA} versions of the images (bottom row). The CycleGAN has adapted the $A$ images to remove their wing-like artefacts and regions of low amplitude underneath the vessels, while preserving much of the original tissue's structure.}
    \label{fig:cycleGAN_adapted_ex_1}
\end{figure*}
\begin{figure*}[]
    \centering
    \includegraphics[width =1.5\columnwidth]{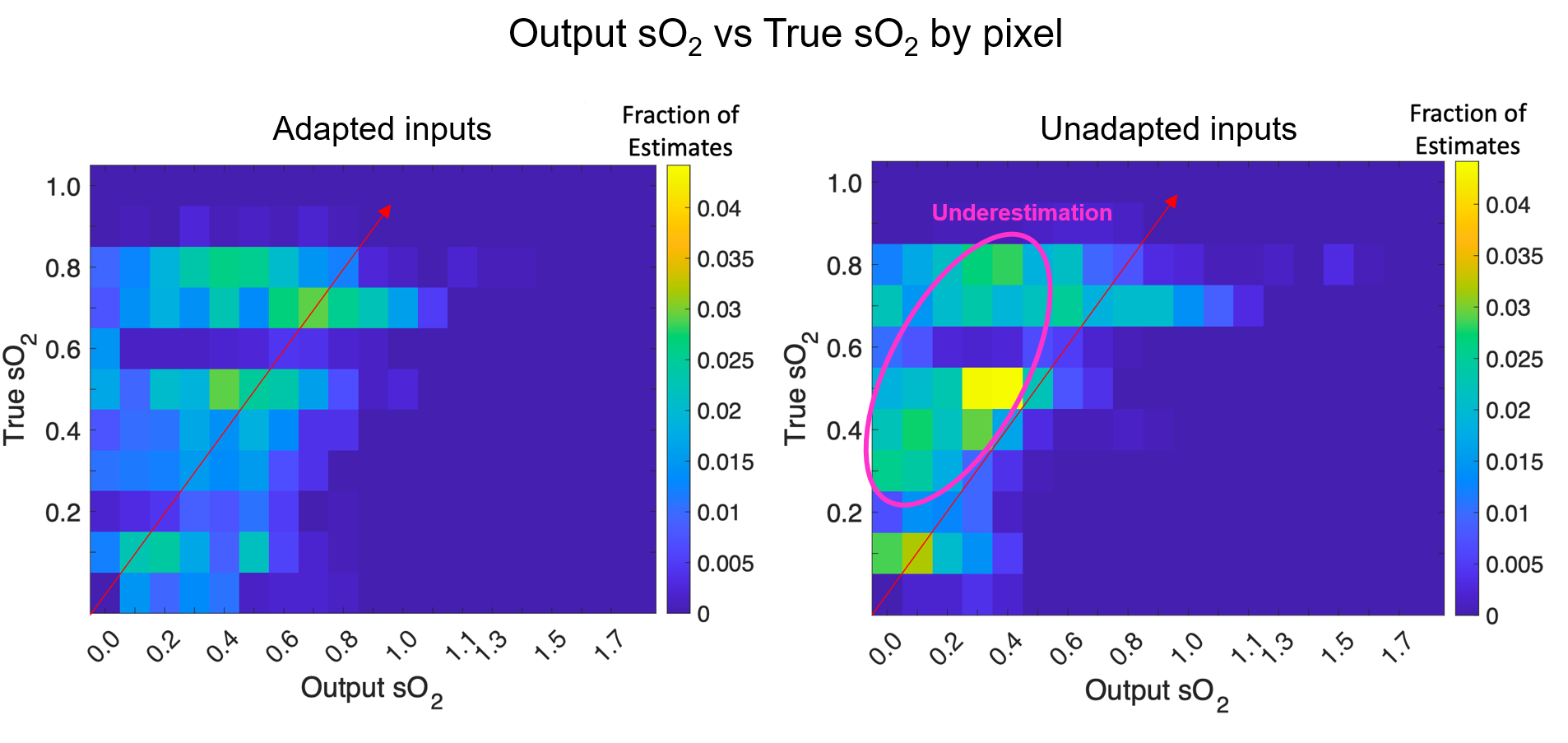}
    \caption{Bivariate histogram showing the accuracy distribution of estimated sO$_2$ values produced from $A$ images adapted to appear \emph{NA} when processed by an sO$_2$-estimating network trained on \emph{NA} images, and when the same unadapted $A$ images were used as inputs to the network. More of the sO$_2$ estimates acquired from the unadapted $A$ inputs are underestimated, while those acquired from the adapted images are spread more evenly around the true sO$_2$.}
    \label{fig:cycleGAN_bivariate_hist}
\end{figure*}
\begin{table*}[]
    \centering
    \begin{tabular}{ |p{4cm}|p{4cm}|p{4cm}|  }
 \hline
 \multicolumn{3}{|c|}{Mean Vascular sO$_2$ estimates} \\
 \hline
 $A$ Images & Adapted Images & Ground Truth \emph{NA} Images\\
 \hline
 30.8\% ($\sigma=$ 31.0\%)  & 39.9\% ($\sigma=$ 30.8\%)  & 49.5\% ($\sigma=$ 27.3\%)\\
 \hline
\end{tabular}
    \caption{The mean of all sO$_2$ estimates within vessels produced from 50 $A$ test images, their adapted counterparts, and the \emph{NA} versions of these test images when used as inputs to an sO$_2$-estimating network pretrained on \emph{NA} images. The mean vascular sO$_2$ estimates produced from the adapted images was 9.1\% closer to the ground truth (percentage points) than those produced from the unadapted images.}
    \label{tab:mean_vess_sO2}
\end{table*}

Once trained, 50 $A$ images not in the training/validation set were adapted using the trained CycleGAN (some examples are shown in Figs. \ref{fig:cycleGAN_adapted_ex_1} and \ref{fig:cycleGAN_adapted_ex_2}). The adapted images and their unadapted counterparts were fed into an sO$_2$-estimating network pretrained on \emph{NA} images (training details given in Appendix \ref{sec:cycleGAN_sO2_train}). The resultant sO$_2$ estimates are summarised in Fig. \ref{fig:cycleGAN_bivariate_hist} and Table \ref{tab:mean_vess_sO2}. 

The mean of all estimated sO$_2$ values from the adapted images (shown in Table \ref{tab:mean_vess_sO2}) was closer to the ground truth than that produced by the unadapted images. For this calculation, the vessels in the original $A$ test images and their adapted counterparts were segmented by setting all pixels in a given image with values less than 0.4 times the max pixel amplitude in the whole image, or less than 0.3 times the max value of all the pixels in their row to zero. Vessel pixels were extracted from the \emph{NA} versions of the images using their known ground truth locations. Observing the bivariate histograms of network outputs in Fig. \ref{fig:cycleGAN_bivariate_hist}, it is evident that fewer sO$_2$ estimates acquired from the adapted images had been underestimated, with a much more equal spread of estimates about the true sO$_2$. These histograms were constructed by only considering non-zero sO$_2$ estimates that had a non-zero corresponding ground truth value. 

\subsection{Generating realistic image data with AmbientGANS}
\begin{figure*}[]
    \centering
    \includegraphics[width = 1.9\columnwidth]{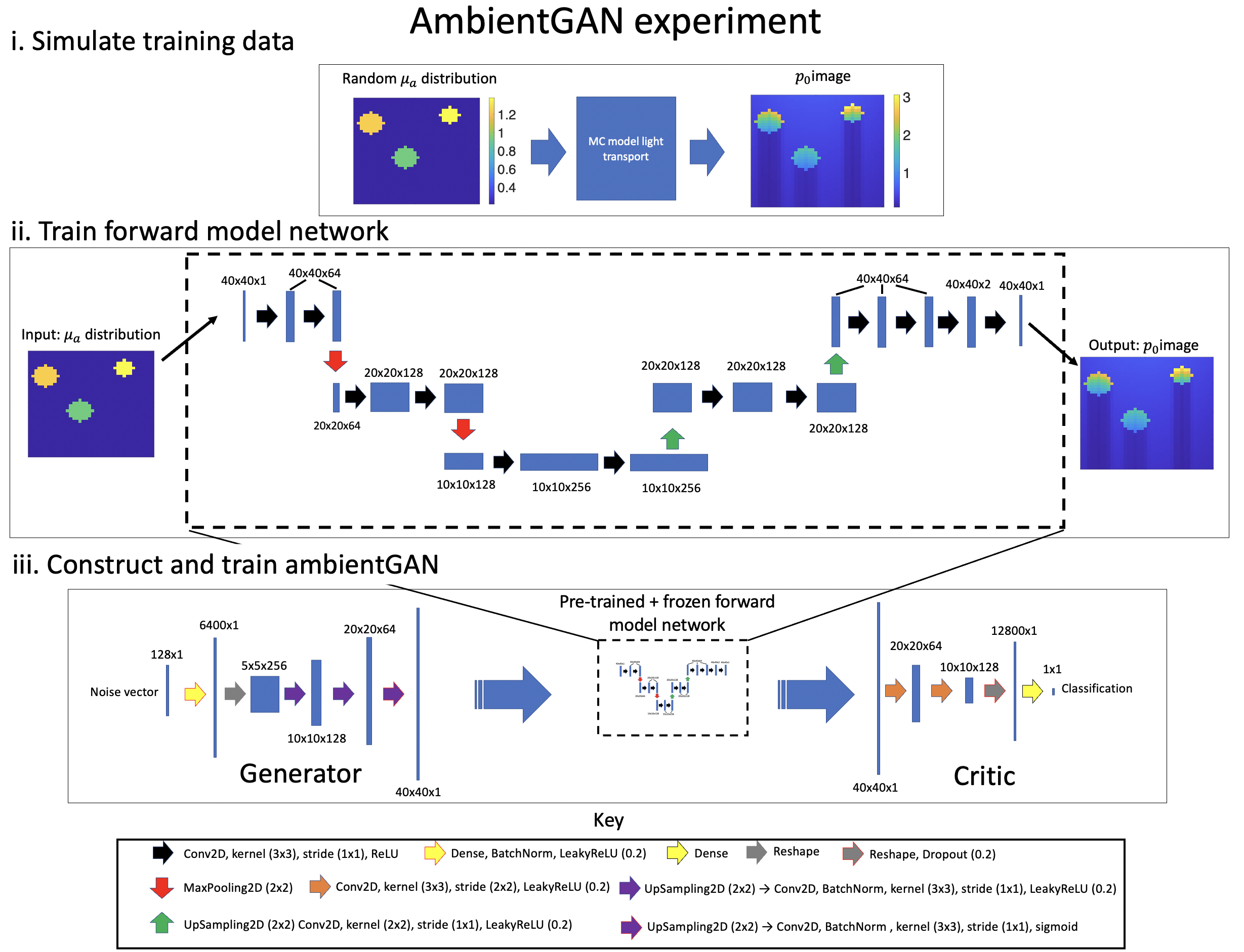}
    \caption{Summary of the AmbientGAN experiment, where we sought to generate simulated $p_0$ images of circular absorbers alongside images of their underlying $\mu_a$ distributions by providing the network with only with $p_0$ images.}
    \label{fig:ambientGAN_summary}
\end{figure*}

\begin{figure*}[]
    \centering
    \includegraphics[width = 1.5\columnwidth]{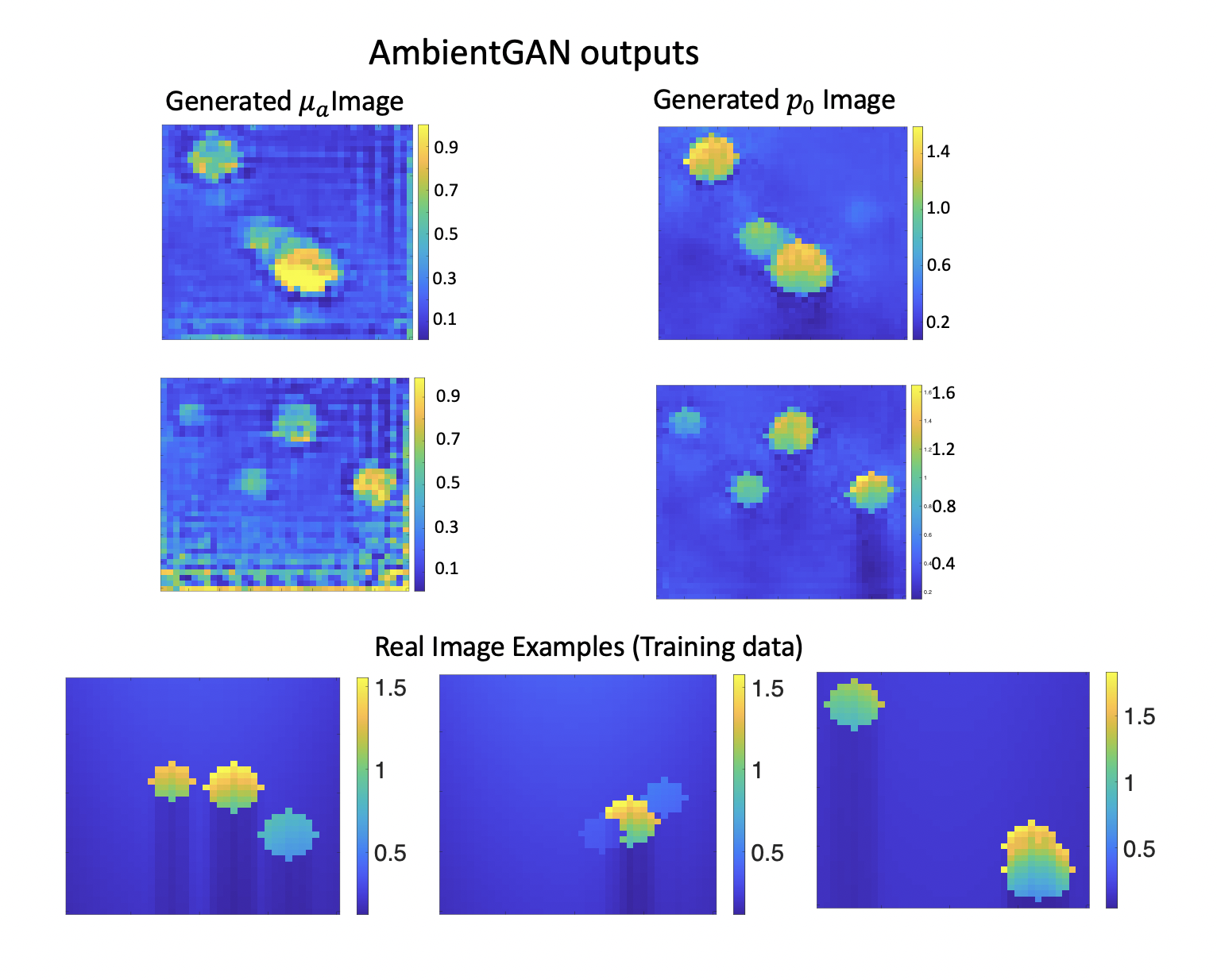}
    \caption{Top: AmbientGAN generated tissue models (images of their $\mu_a$ distribution) and their corresponding generated $p_0$ images. Bottom: A few non-generated $p_0$ images fed to the AmbientGAN's critic.}
    \label{fig:ambientGAN_results}
\end{figure*}
We sought to provide a basic demonstration of how AmbientGANs could be used to generate paired training data for learning chromophore quantification. We constructed a WGAN-GP with a frozen and pretrained network approximating a light propagation model (that outputs an image of $\mu_a\Phi$) appended to the generator. The architecture is shown in Fig. \ref{fig:ambientGAN_summary}. Therefore, the generator was trained to generate tissue models that, when fed through the light model network, would produce images of $\mu_a\Phi$ (referred to here as $p_0$ images) indistinguishable from those presented to the critic. The critic (discriminator) was fed 2D PA images ($40\times40$ pixels) (i.e. $p_0({x},\lambda) = \Gamma\mu_a({x},\lambda)\Phi({x},\lambda))$, where ${x}$ is each pixel in the image, $\lambda$ is the excitation wavelength, and $\Gamma=1$ is the PA efficiency) of circular absorbers immersed in a homogeneous absorbing background medium. Each tissue model contained between 1-3 circular absorbers with radii between 3 and 5 pixels. The homogeneous absorbing background was assigned an absorption coefficient between 0 and .5 mm$^{-1}$. Each circular absorber was assigned a random absorption coefficient computed by adding a random number between 0 and 1.3 mm$^{-1}$ to the background's absorption coefficient (this ensured each absorber had a higher absorption coefficient than the background, as is generally the case with vessels immersed in tissue). Fluence simulations were run in MCXLAB \cite{fang_2011}. The tissue was illuminated using a truncated, collimated Gaussian beam centred at pixel 20 with a waist-radius of 50 pixels. The pixel length was set to have real dimensions of 0.1 mm, and the refractive index was homogeneous throughout the tissue with a value of 1.4. The timestep was set to $10^{-11}$s with a total time of $10^{-9}$s. A total of $10^{7}$ photons were used for each simulation.

The light model network (architecture shown in Fig. \ref{fig:ambientGAN_summary}). was trained on 500 image pairs (an image of the tissue's absorption coefficient distribution as an input and the corresponding image of $\mu_a\Phi$ as the ground truth) for 80 epochs. A validation set of 100 examples was used to determine the stopping point. The network was trained with a loss function of the Euclidean norm of the squared difference between the predicted image and the ground truth, along with Adam as the optimisation algorithm.

The AmbientGAN was trained with 399 $p_0$ images (simulated in the same way as discussed above) with the same loss functions, critic/generator architectures, and hyperparameters used in the WGAN-GP example shown in Appendix \ref{sec:Appendix_WGAN}) for 5000 epochs. The outputs shown indicate that the AmbientGAN has produced plausible underlying absorption distributions for the generated images (e.g. absorbers with high intensity at depth have correspondingly high absorption coefficient values). Though, it failed to reproduce the shadowing effect in the generated $p_0$ images (both within the vessels and underneath them), and homogeneous $\mu_a$ estimates within the vessels. The generated absorption coefficient images contain fluctuating values at larger depths. This likely occurs because the absorption coefficient in these regions do not have a large effect on the final image amplitude due to the decay in the fluence. Therefore, the network has no incentive to keep absorption values at greater depths constant and the same as the rest of the background. This phenomenon has been found in the results of other inversion schemes, such as \cite{cox2006quantitative,cox2006two,cox2005quantitative}. The generated tissue models also contain unrealistic tile/checkerboard artefacts. Such artefacts have been reported to emerge as a consequence of choices in stride/kernel size, though this has not been investigated here \cite{aitken2017checkerboard}. 

\section{Discussion}

We have described two GAN-based strategies for reducing the domain gap between synthetic PA images used for training and real PA images used as test data. With a toy example, we have shown that unsupervised domain adaptation performed using a CycleGAN can align the feature spaces of images belonging to two different data domains. The adaptation of the target input data ($A$ images) so that its feature space is more closely aligned with that of the source domain (\emph{NA} images) can improve the accuracy of the mean of their sO$_2$ estimates produced by an sO$_2$-estimating network trained on data from the source domain. However, even in the idealised case presented here, the approach was only capable of `shifting' sO$_2$ estimates to be more equally distributed about the ground truth, and did not appear to reduce the variance in the accuracy of estimates. It is clear that the constraints on the adaptation process are inadequate. An ideal adaptation of an $A$ image would produce an image identical to one that would be simulated if the forward model used to simulate \emph{NA} images was applied to the same underlying tissue model. However, no such constraint was applied, and it is not clear how one could be formulated. Another limitation of this technique is that unless there are ground truths available from both the source and target domains, its use is practically limited to cases where the output feature spaces of both domains are similar.

We have also illustrated how the AmbientGAN framework can be used to generate paired training data for sO$_2$-estimating/chromophore quantification networks, only providing it PA images for training. However, there are two key challenges with the approach. The first is formulating a differentiable forward model of image generation that is computationally efficient to compute. This was bypassed in this study by training a network to approximate the forward model. However, in practice, a paired dataset of the optical properties of a series of tissues and their corresponding PA images will not be available. Secondly, even in the case where an efficient and differentiable forward model can be appended to the generator, there is no clear way to constrain it to ensure the generated tissue models are indeed representative of real tissues without the use of prior knowledge of how optical properties are typically distributed \emph{in vivo}. The same limitation applies to recent work described in \cite{dreher2023unsupervised}, where conditional generation is used to preserve the spectral features of simulated images adapted to appear more realistic. Indeed, outside of the techniques proposed here, a wide range of generative modelling frameworks could be used to learn the adaptation of network inputs (e.g. improving the quality of adaptations through the use of auxiliary networks \cite{gong2022ssast}, or more state of the art frameworks such as diffusion models \cite{chen2022generative, yang2022diffusion, cao2022survey}). But despite possible improvements to adapted image quality, it is not immediately clear how any of these frameworks could improve the quality of the ground truth concentration distributions.

With that said, in the event where this information can be drawn from a set of tissue examples, then it is not unreasonable to suggest that generative deep learning could accelerate the generation of large amounts of training data. Aside from the applications proposed here, adaptive models could also be used for other tasks related to chromophore quantification. For example, it might be possible to use the CycleGAN framework for approximation error modelling \cite{tarvainen2016image, tick2019modelling}. I.e. when adapting \emph{NA} images to appear $A$, the CycleGAN can be thought of as implicitly learning how i) the inclusion of the acoustic and reconstruction models affect image properties, and ii) how to execute a post-processing step that incorporates them into the outputs of the light model that produces the \emph{NA} images.

It is evident that generative models are versatile tools and that this work has by no means provided a complete discussion of their possible uses for chromophore quantification or photoacoustic imaging more generally. Further investigation is needed to understand the true impact these techniques could have on realising a learned model for \emph{in vivo} chromophore quantification.

\section*{Acknowledgments}
The authors would like to thank Andreas Hauptmann, Antonio Stanziola, and Simon Arridge for useful discussions. CB acknowledges funding from the London Interdisciplinary Doctoral Training Programme (LIDo). 
\begin{figure}[!h]
    \centering
    \includegraphics[width=.9\columnwidth]{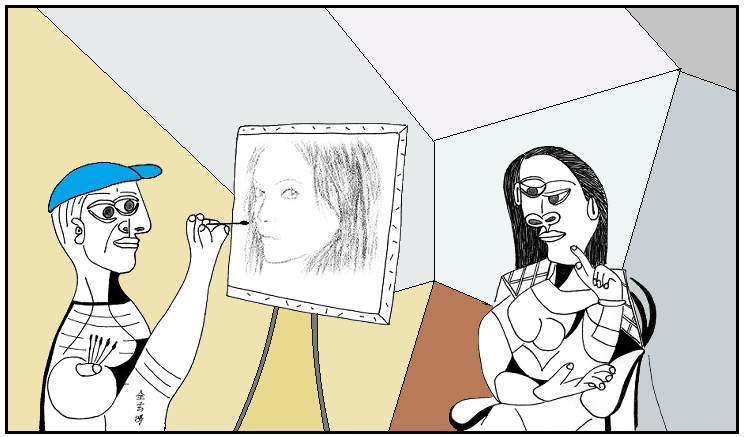}
    \caption{\cite{abstruse_goose}}
    \label{fig:picasso}
\end{figure}

\bibliography{bibliography.bib}
\onecolumn
\newpage
\appendix
\subsection{WGAN-GP}
\label{sec:Appendix_WGAN}
\begin{figure*}
    \centering
    \includegraphics[width=.85\columnwidth]{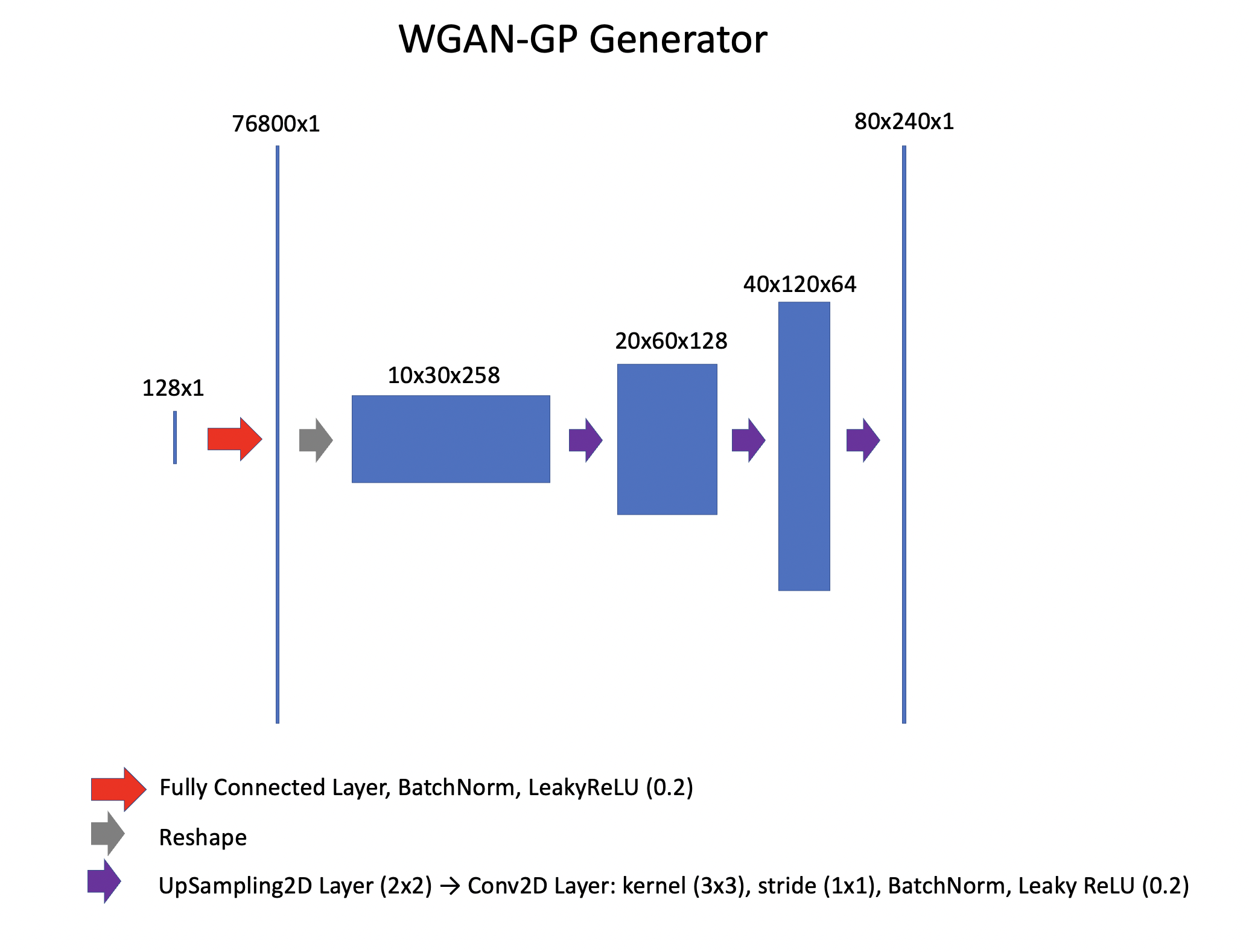}
    \includegraphics[width=.7\columnwidth]{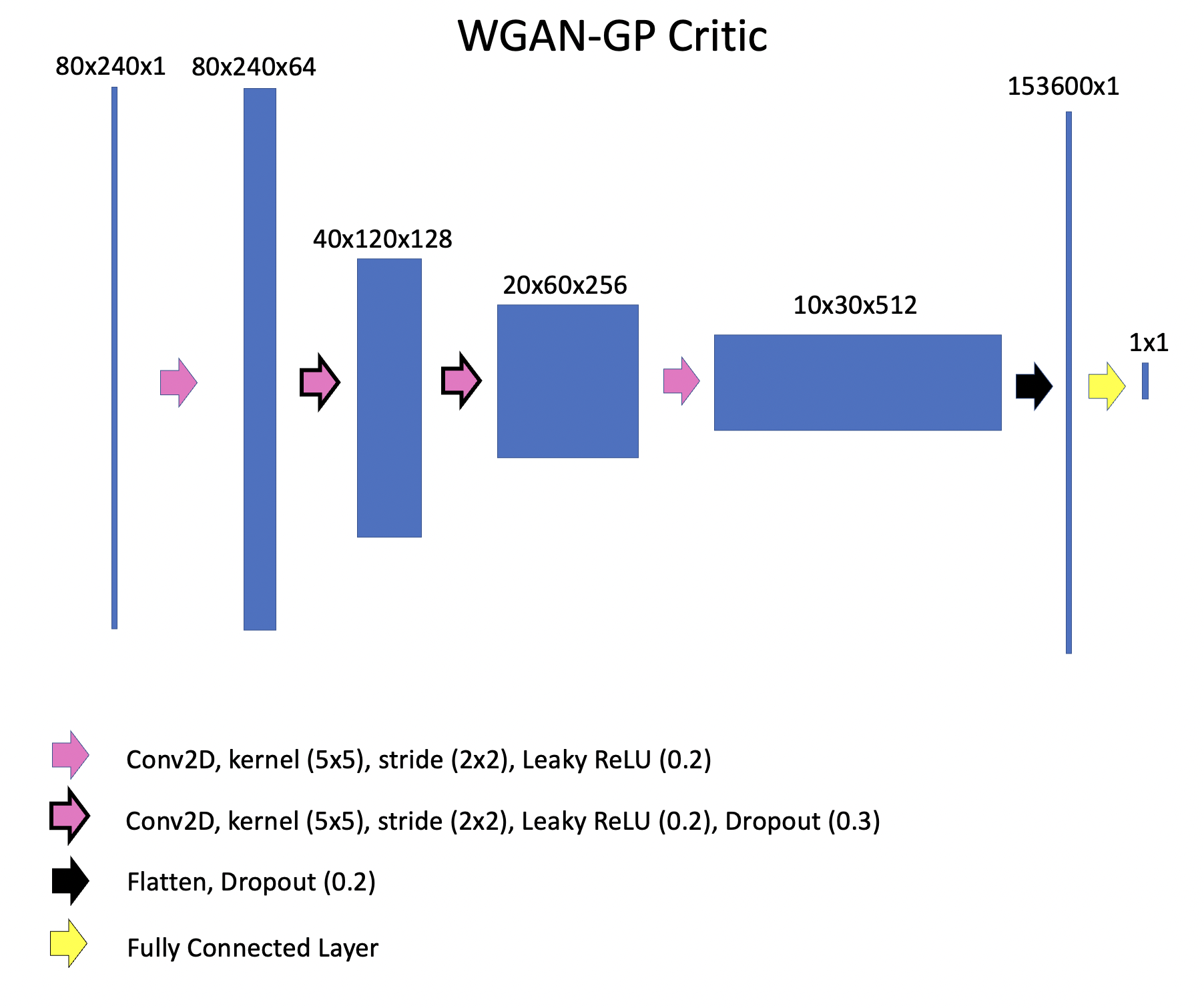}
    \caption{Architecture for the WGAN-GP used to generate synthetic PA images of human forearms.}
    \label{fig:WGAN_arch}
\end{figure*}

We trained a Wasserstein GAN with Gradient Penalty (WGAN-GP) (architecture shown in Fig. \ref{fig:WGAN_arch}) to generate synthetic PA images of human forearm vessels. The network was trained on a dataset of reconstructed images regularised with total variation as described in \cite{hauptmann2018model}. Specifically, the 21 3D images were parsed into a 504 image 2D training set by taking a slice every 20 pixels along both axes where the skin layer is at the top. Code from \cite{Nain2020} was adapted for our task.

\begin{figure*}
    \centering
    \includegraphics[width=.8\columnwidth]{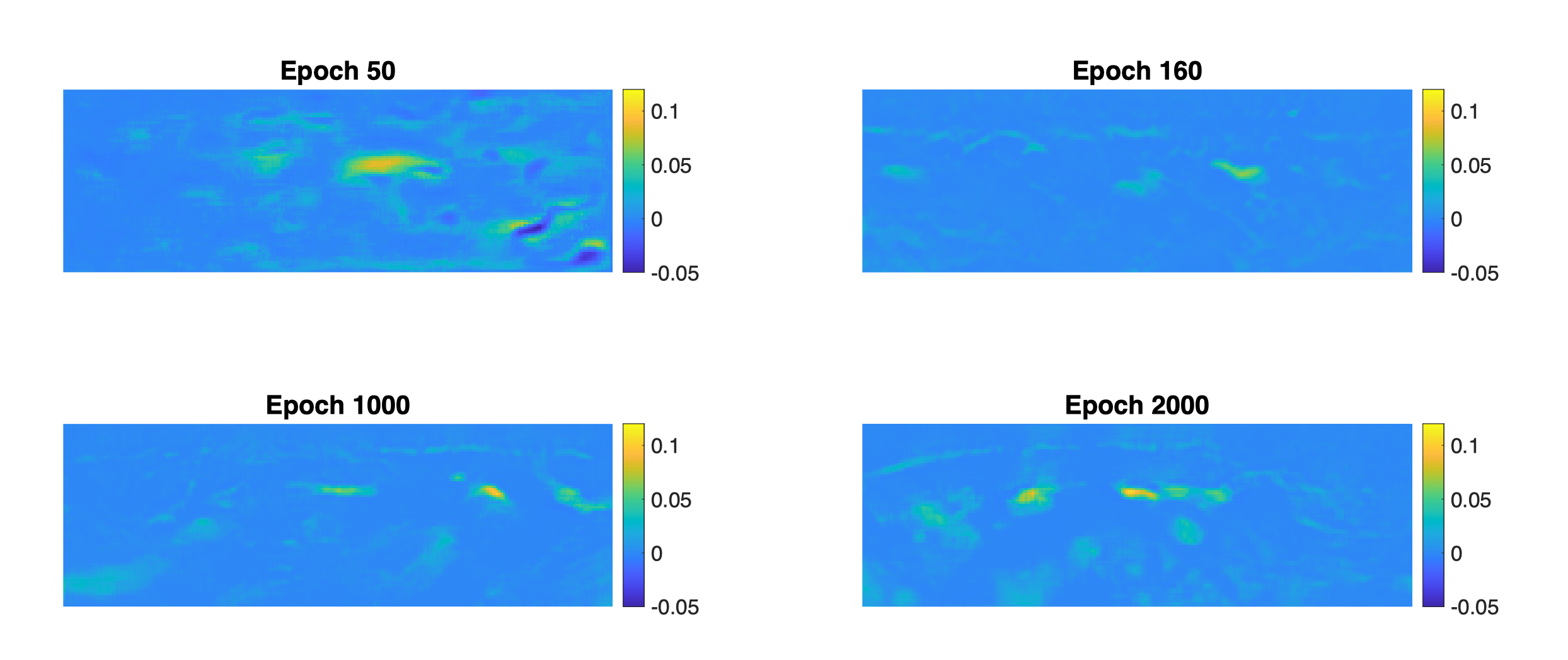}
    \caption{WGAN-GP outputs at various epochs. We can see the quality of the generated human forearm images improves considerably from epoch 50.}
    \label{fig:WGAN_outputs_epoch}
\end{figure*}

\begin{landscape}
\begin{figure*}
    \centering
    \includegraphics[width=.99\columnwidth]{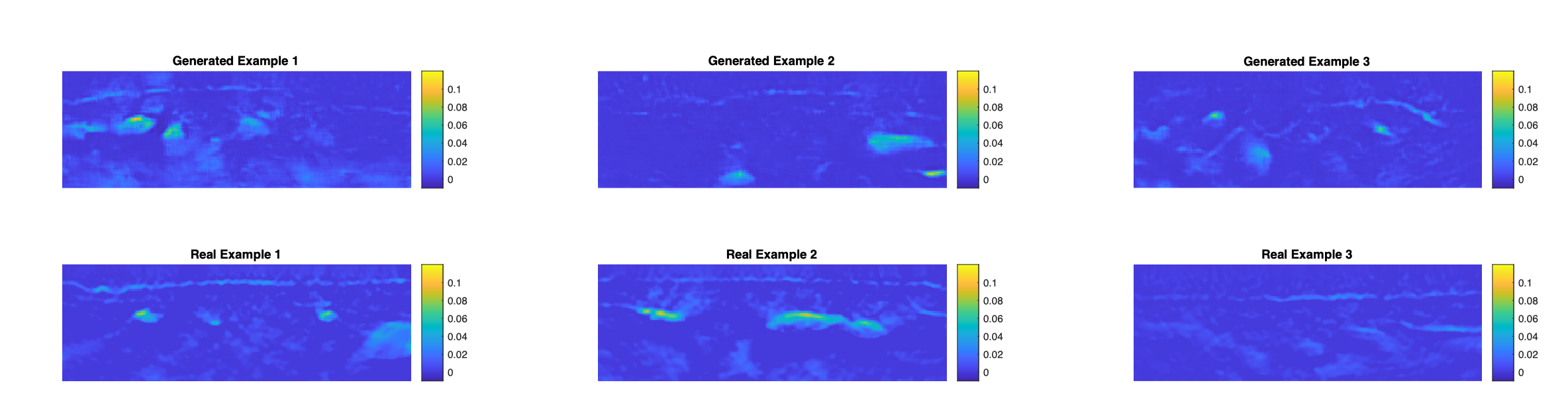}
    \caption{WGAN-GP generated PA images of human forearms produced after 2000 epochs. Generated images feature a superficial skin layer with vessels immersed underneath. Various background absorbers are scattered throughout the tissue.}
    \label{fig:WGAN_outputs}
\end{figure*}
\end{landscape}

All modules were trained with Adam as the optimizer, learning rates of 0.0002, $\beta_1$ = 0.5, $\beta_2$ = 0.9 (the exponential decay rates for the first and second moment estimates respectively), and batch sizes of 10. The training loop was constructed to update the generator after every 3 critic updates (a hyperparameter for the WGAN-GP). The data was normalised by subtracting the mean of the combined training and validations sets before putting them through the critic. Noise vectors of length 128 sampled from a normal distribution were used as inputs to the generator, which was trained for a total of 2000 epochs (loss curves are shown in Fig. \ref{fig:cycleGAN_losses}, where the critic loss was calculated at the time of each generator update). Training was terminated once the generator appeared to produce examples realistic enough to convince a casual human observer (therefore reaching a suitable standard of quality for our proof of concept example). Some example outputs are shown in Figs. \ref{fig:WGAN_outputs} and \ref{fig:WGAN_outputs_epoch}. The network has learned to output a distinct skin layer above vessel-like structures immersed in an arrangement of randomly distributed background absorbers.

The objective function for the WGAN-GP is given by, 
\begin{equation}
    L = \mathbb{E}_{{\tilde{x}}\sim \mathbb{P}_g}[D({\tilde{x}})] - \mathbb{E}_{{x}\sim \mathbb{P}_r}[D({x})] + \mathbb{E}_{{\hat{x}}\sim \mathbb{P}_{\hat x}}[(|| \nabla _{\hat x}D({\hat x})||_2 -1)^2].
    \label{eq:WGAN_GP_loss}
\end{equation}

In \cite{gulrajani2017improved}, $P_{\hat x}$ is defined as ``sampling uniformly along straight lines between pairs of points sampled from the data distribution $P_r$ and the generator distribution $P_g$". The gradient penalty term is acquired using images interpolated from pairs of real and generated examples. This step is performed to make the calculation of the gradient penalty term more computationally tractable \cite{foster2019generative}. The training process is executed in the following manner:

\begin{enumerate}
    \item Real images $x\sim P_r$, noise $z\sim P_z$, and a random number from a uniform distribution $\epsilon\sim U[0, 1]$ are sampled.
    \item The noise is used as an input to the generator to produce synthetic images: $\tilde{x}\leftarrow G_{\theta}(z)$.
    \item Interpolated images $\hat{x} \leftarrow \epsilon{x} + (1 - \epsilon)\tilde{{x}}$ are acquired and the gradient penalty term is calculated.
    \item The real and fake images are input to the critic.
    \item The critic outputs from step 4 are used to compute the loss $L$, and the critic is updated using gradient descent, minimising $L$. 
    \item Steps 1-5 are performed $n$ times, where $n$ is a hyperparameter.
    \item Additional noise vectors are sampled $z\sim P_z$, and used to generate additional synthetic images.
    \item These synthetic images are input to the critic. The mean score $\mathbb{E}_{{\tilde{x}}\sim \mathbb{P}_g}[D({\tilde{x}})]$ is then calculated. 
    \item The generator parameters are updated by performing gradient descent, minimising $-\mathbb{E}_{{\tilde{x}}\sim \mathbb{P}_g}[D({\tilde{x}})]$
\end{enumerate}

\begin{figure}
    \centering
    \includegraphics[width=.5\columnwidth]{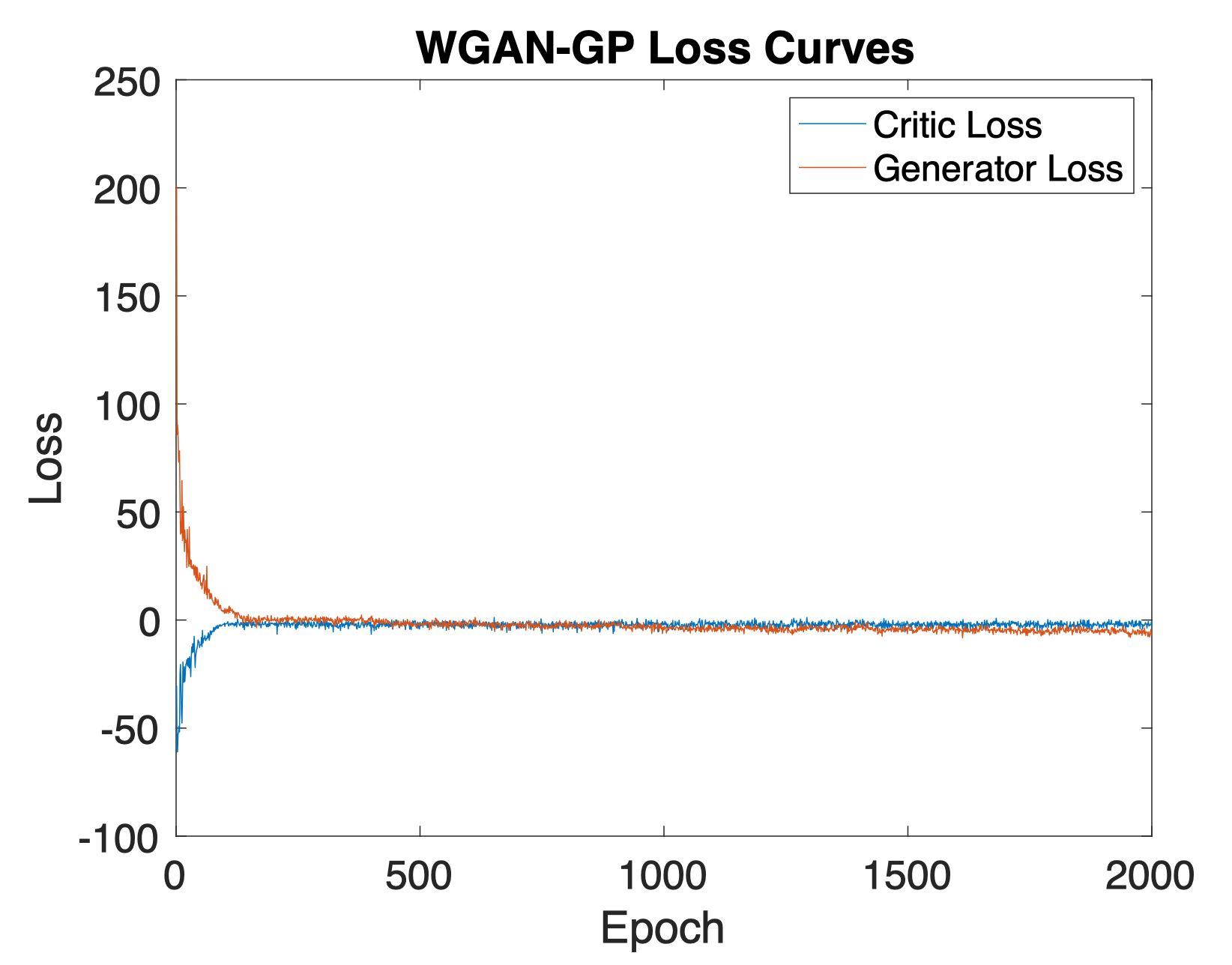}
    \caption{Loss curves for training the WGAN-GP to generated synthetic PA images of human forearms.}
    \label{fig:WGAN_losses}
\end{figure}

\newpage
\subsection{CycleGAN experiment: evidence of domain-mismatch between $A$ and NA images}
\begin{figure*}[h!]
    \centering
    \includegraphics[width=.45\columnwidth]{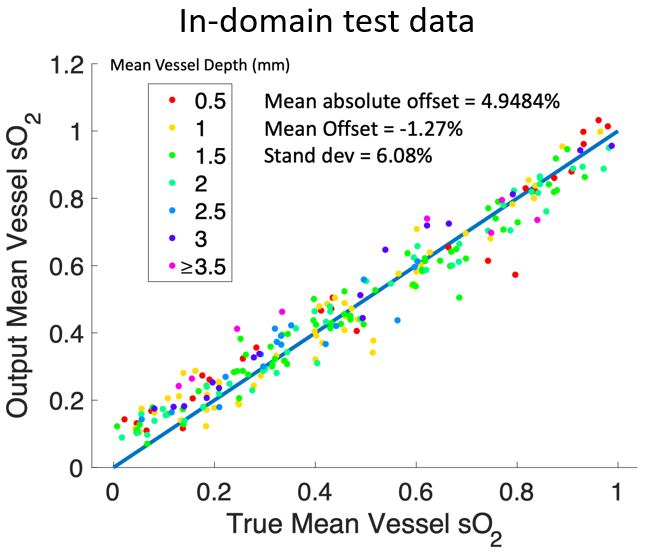}
    \includegraphics[width=.45\columnwidth]{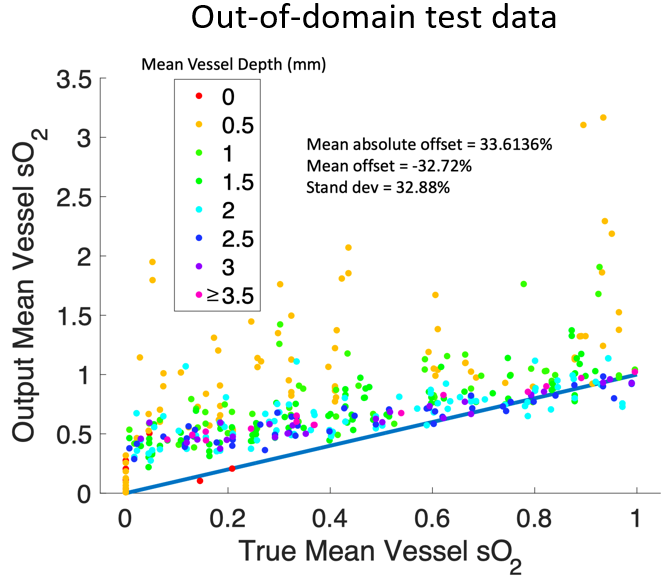}
    \caption{Left: Plot of mean vascular sO$_2$ values produced by inputting 50 $A$ 3D images into a network pretrained on $A$ images. Right: The same kind of plot, but produced when 50 \emph{NA} test images were processed by the network. The difference in accuracy shows that $A$ and \emph{NA} images belong to different data domains.}
    \label{fig:mean_vess_sO2_generalisability}
\end{figure*}

\begin{figure*}[h!]
    \centering
    \includegraphics[width=.9\columnwidth]{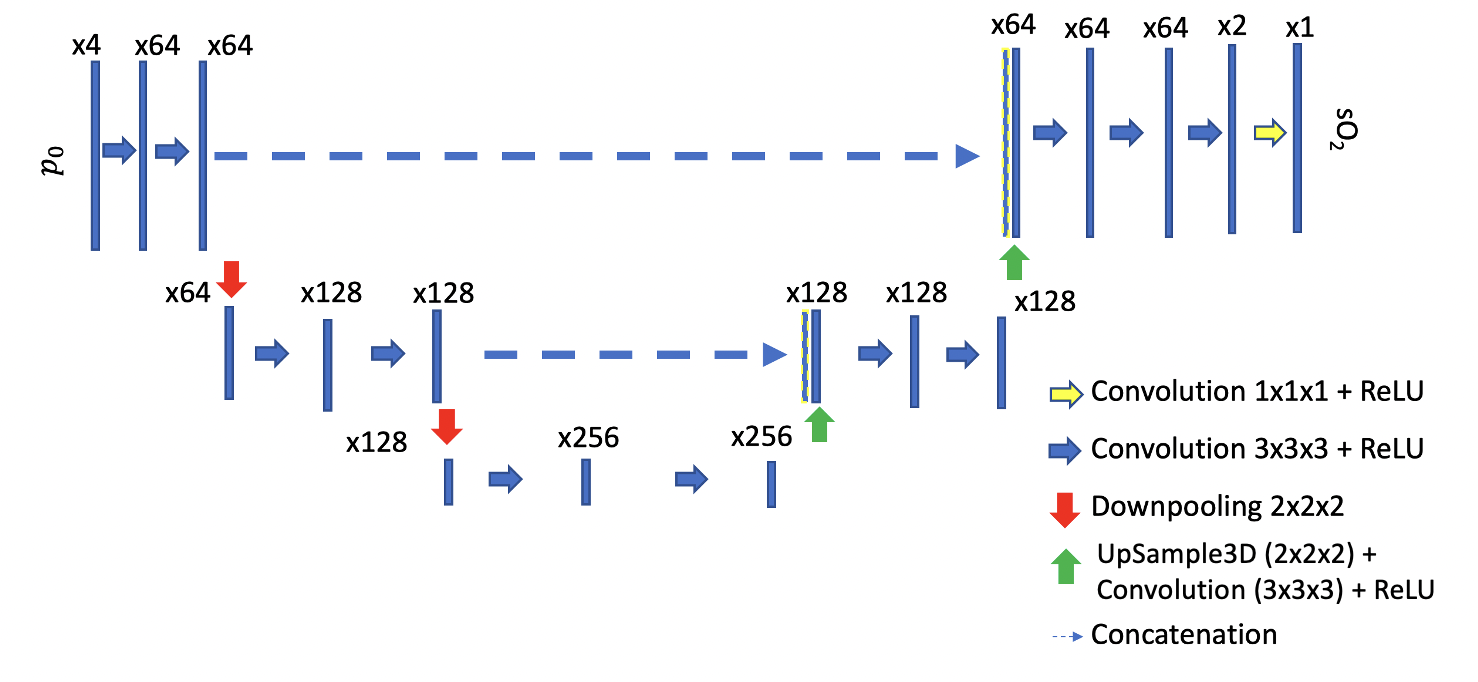}
    \caption{Architecture of the sO$_2$ network used to assess the domain mismatch between $A$ and \emph{NA} images. The segmentation network also used for this task had the same architecture.}
    \label{fig:generalisability_arch}
\end{figure*}

Here show the existence of a domain gap between the $A$ and \emph{NA} images used in the CycleGAN domain adaptation experiment. Fig. \ref{fig:mean_vess_sO2_generalisability} shows the accuracy of mean vessel sO$_2$ estimates produced from 50 3D test images fed to an sO$_2$-estimating network (architecture shown in Fig. \ref{fig:generalisability_arch}) trained on $A$ images (training details and the vessel segmentation procedure are provided below). The figure also shows that estimates produced from 50 \emph{NA} test images were much less accurate. If $A$ and \emph{NA} images were described by the same domain, the accuracy of the estimates acquired from either set should be more similar. A histogram of all estimates produced from either set show a similar discrepancy.

\subsubsection{sO$_2$-estimating network for assessing domain mismatch}
\label{sec:sO2_net_domain_gap_R_NR}
The sO$_2$-estimating network was trained on 400 images for 25 epochs, using a validation set of 50 images, a batch size of 5, a 2-Norm loss function, with Adam as the optimisation algorithm, and an initial learning rate of $10^{-4}$. A segmentation network with an identical architecture and hyperparameters was trained on 400 images for 15 epochs, where the ground truth was instead a 3D image of the tissue's vessels. The inputs for each network were images of a tissue simulated with optical illumination wavelengths of 784 nm, 796 nm, 808 nm, and 820 nm. 

Mean vessel sO$_2$-estimates were calculated in the following manner. First, the indices associated with each major body in the segmentation network output were identified. This was achieved by thresholding the output of the segmentation network $S$ so all voxels with intensities $<0.2$ were set to zero, producing a new image $S'$. This was done to remove small values which connected all the vessels into one large body, ensuring each vessel was isolated in the volume. The output values from the segmentation network are approximately in the range 0-1 because the segmented training data images were binary, so the threshold of 0.2 (chosen empirically) was applicable to all the output images without requiring an additional normalisation. Then, the indices associated with each major body in $S'$ were identified using the \verb|bwlabeln()| MATLAB function, generating a labelled image $L$, where each body in the image was assigned a unique value to be identified by, and all the voxels in a given body share the same value.

Then, $S'$ was thresholded so all voxels with intensities $<0.8$ were set to zero, producing a new image $S''$. This was done to isolate voxels where the network was confident that vessels were present.

All the voxels in $L$ that have corresponding values of zero in $S''$ were also set to zero, producing a new image $L'$. Voxels that once belonged to an unbroken independent body before this thresholding step may now be in separate bodies. However, their voxel ID retains information about the body to which they originally belonged. 

The mean sO$_2$ of the voxels sharing the same integer value in $L'$ (i.e. those belonging to the same original vessel) were calculated using the corresponding values in the output of the sO$_2$ estimating network. The ground truth mean sO$_2$ of the voxels sharing the same integer value in $L'$ were calculated using the values from the ground truth sO$_2$ distribution.

\label{sec:Appendix_poor_generalisation}

\newpage
\subsection{CycleGAN experiment: CycleGAN training details}
\label{sec:cycleGAN_train}
\begin{figure*}
    \centering
    \includegraphics[width = .9\columnwidth]{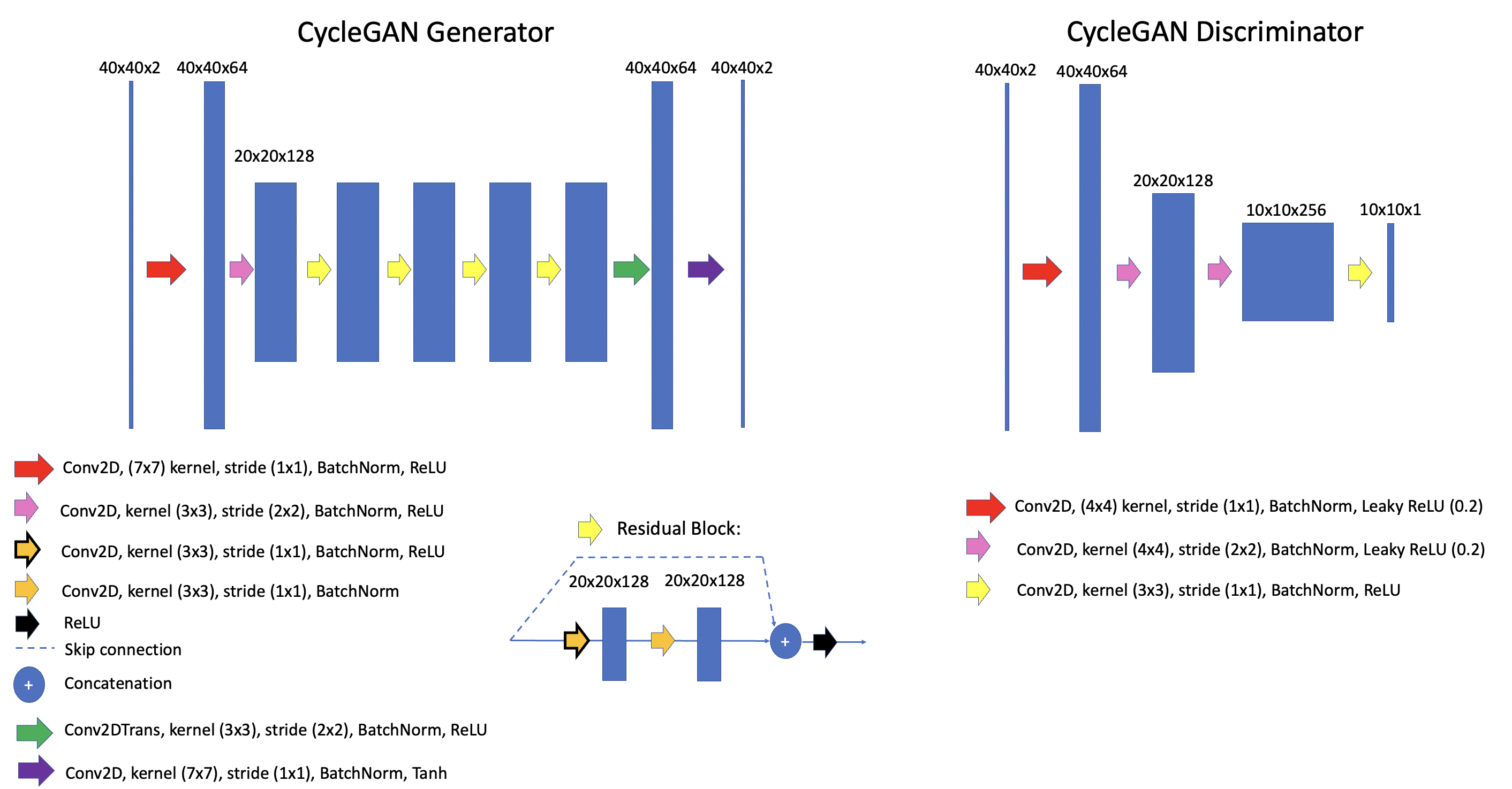}
    \caption{Architectures used for the generators (left) and discriminators (right) in the CycleGAN. The generator is based on the ResNet, which features short-range skip connections to alleviate model degradation with increasing layer size. }
    \label{fig:cycleGAN_arch}
\end{figure*}

\begin{figure*}[h!]
    \centering
    \includegraphics[width = .7\columnwidth]{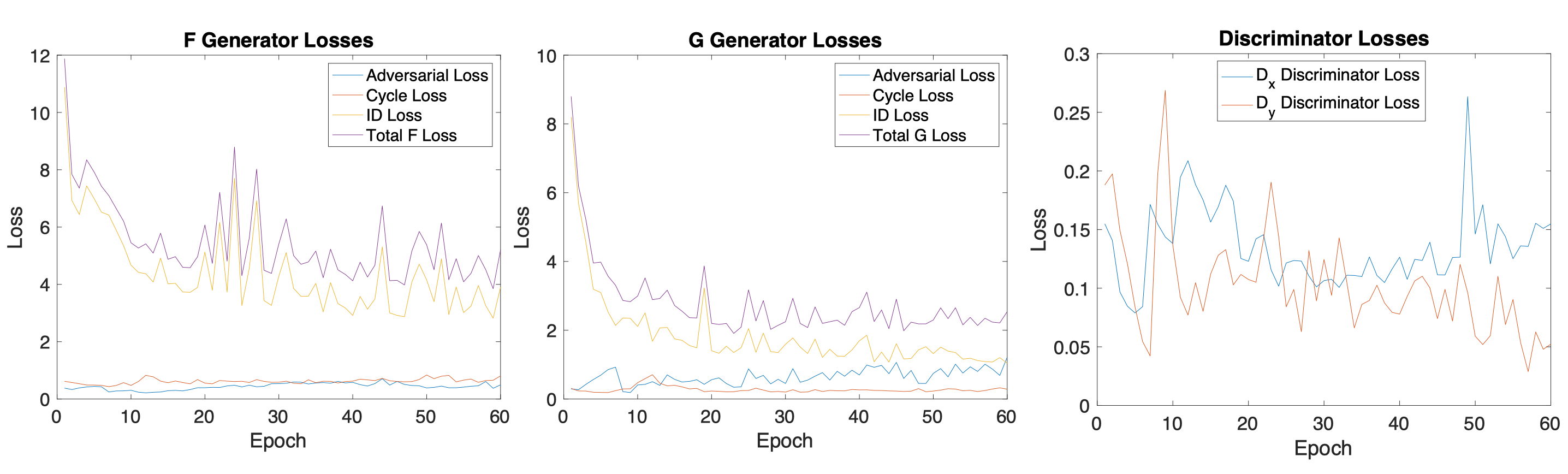}
    \caption{Loss curves for the CycleGAN generators and discriminators. $G$ refers to the generator that adapts $A$ images into \emph{NA} images, while $F$ refers to the generator that adapts \emph{NA} images into $A$ images. Discriminator $D_x$ refers to the discriminator network that discerns whether an image is a true $A$ image or one that has been adapted to appear like an $A$ image, while $D_y$ refers to the discriminator that discerns whether an image is a true \emph{NA} images or one that has been adapted to appear like an \emph{NA} image.}
    \label{fig:cycleGAN_losses}
\end{figure*}

Here we provide further details about the CycleGAN's training procedure. Each step of the training process is summarised below. Loss curves are provided in Fig. \ref{fig:cycleGAN_losses}. The network architecture is provided in Fig. \ref{fig:cycleGAN_arch}.

\begin{enumerate}
    \item The $A$ and \emph{NA} images are normalised by subtracting the mean calculated from all 600 examples in their respective datasets, and dividing by the difference between their respective maximum and the minimum values. 
    \item 400 images from each set were set aside for training. A batch of images $x$ and $y$ from each domain $X$ ($A$ images) and $Y$ (\emph{NA} images) are extracted.
    \item Synthetic images $G(x)$ and $F(y)$ are generated from these batches.
    \item $F(G(x))$ and $G(F(y))$ are acquired to compute the cycle consistency loss.
    \item $F(x)$ and $G(y)$ are acquired to compute the identity loss.
    \item The generated and original images are then fed to their respective discriminators: $D_y(G(x))$, $D_x(F(y))$, $D_y(y)$, $D_x(x)$. 
    \item{The losses are then calculated:}
\begin{itemize}
    \item Generator Loss (adversarial + cycle + identity)
    \begin{itemize}
        \item Adversarial (Mean Square Error - MSE)
        \begin{itemize}
            \item MSE$({A,D_x(F(y))})$
        \item MSE$({A,D_y(G(x))})$
        \item($A$ is a matrix of ones, with the same dimensions as the outputs of $D_x$  and $D_y$)
        \end{itemize}
        \item Cycle (Mean Absolute Error - MAE)
        \begin{itemize}
            \item $\lambda_{C}$MAE$(x,F(G(x)))$
            \item $\lambda_{C}$MAE$(y,G(F(y)))$
            \item where $\lambda_{C}=100$
        \end{itemize}
        \item Identity
        \begin{itemize}
            \item $\lambda_{ID}$$||(x - F(x))||_{2}$
            \item $\lambda_{ID}$$||(y - G(y))||_{2}$
            \item where $\lambda_{ID}=25$
        \end{itemize}
    \end{itemize}
    \item Discriminator Losses
        \begin{itemize}
            \item $\sqrt{\mbox{MSE}(B_{1}, D_x(x)) + \mbox{MSE}(B_{0}, D_x(F(y)))}$ 
            \item $\sqrt{\mbox{MSE}(B_{1}, D_y(y)) + \mbox{MSE}(B_{0}, D_y(G(x)))}$ 
            \item $B_{1}$ is a matrix of ones, with the same dimensions as outputs of $D_x$ and $D_y$ 
            \item $B_{0}$ is a matrix of zeros, with the same dimensions as outputs of $D_x$ and $D_y$ 
        \end{itemize}
        \item Then, the generators and discriminators are updated:
            \begin{itemize}
                \item Adam used as the optimizer for all networks with a learning rate of $0.0002$, an exponential decay rate for the 1st moment estimates of $\beta_{1} = 0.5$, and the exponential decay rate for the 2nd moment estimates of $\beta_{2}=0.999$. 
            \end{itemize}
\end{itemize}
\end{enumerate}

\newpage
\subsection{CycleGAN experiment: sO$_2$-net training procedure}
\begin{figure*}[htp!]
    \centering
    \includegraphics[width = .8\columnwidth]{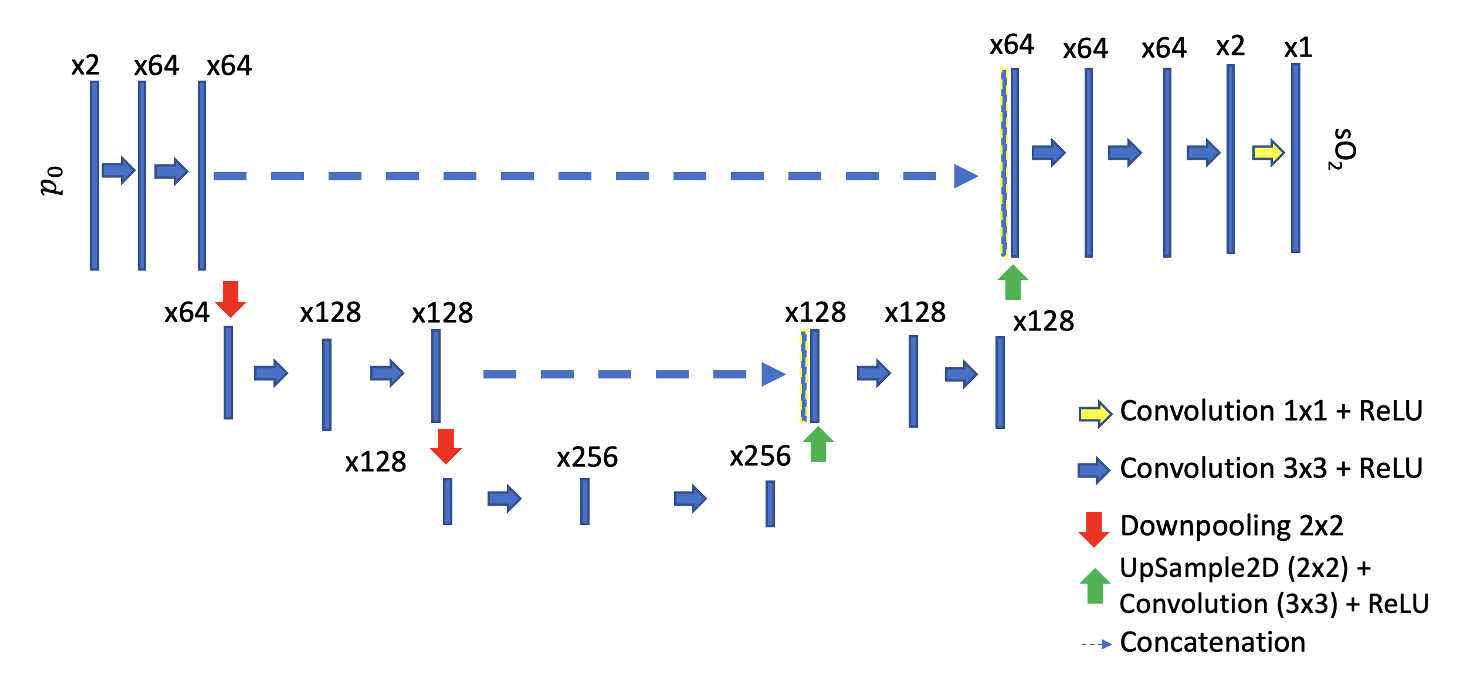}
    \includegraphics[width = .5\columnwidth]{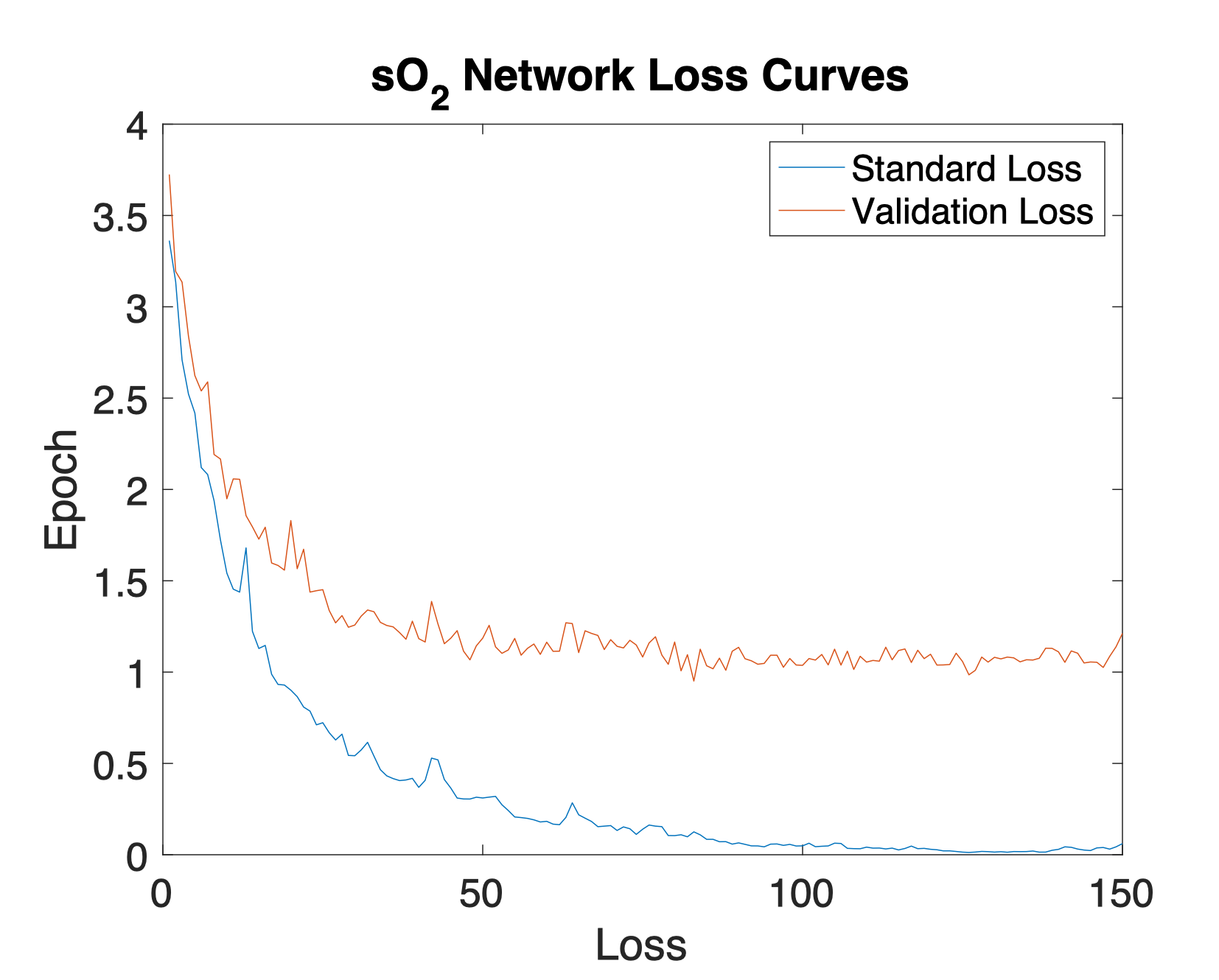}
    \caption{Top: Network architecture of the sO$_2$-net used to process both reconstructed images and their adapted counterparts in the unsupervised domain adaptation/CycleGAN experiment. Bottom: Loss curves for the sO$_2$-net. The network's parameters at 100 epochs was used for evaluation.}
    \label{fig:sO2_net_arch_UDA}
\end{figure*}

\begin{figure*}[htp!]
    \centering
    \includegraphics[width = .45\columnwidth]{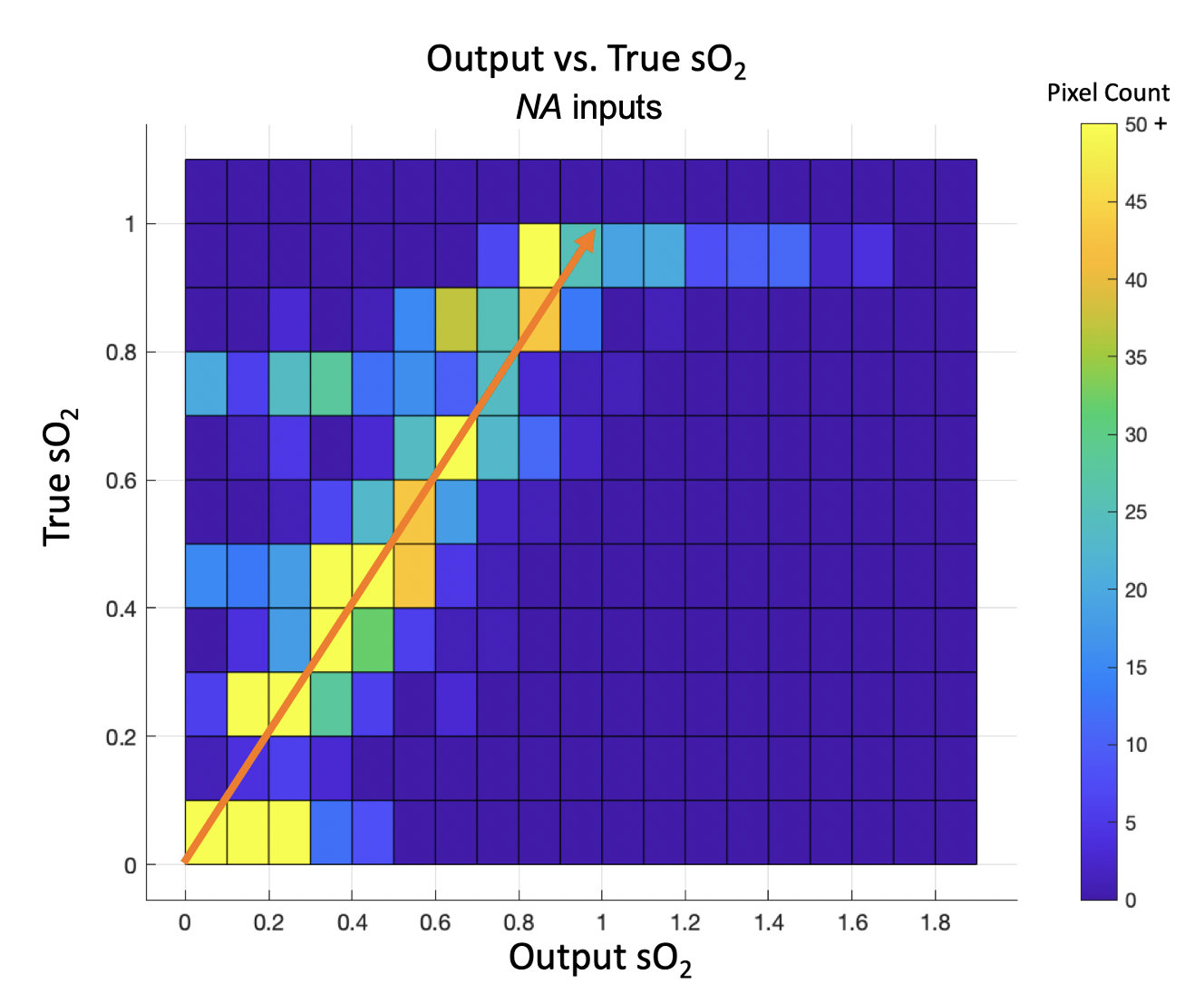}
    \caption{Histogram showing the distribution of sO$_2$ estimates produced by the sO$_2$-estimating network pretrained on \emph{NA} images when provided in-domain test data as inputs. This demonstrates the accuracy of the pretrained sO$_2$-estimating network used in the CycleGAN experiment.}
    \label{fig:UDA_sO2_net_evaluation}
\end{figure*}

The CycleGAN domain adaptation experiment featured the use of a pre-trained sO$_2$-estimating network to process adapted and unadapted $A$ images. Here we provide details about how this network was trained and about its architecture. The sO$_2$-estimating network was a 2D encoder decoder with skip connections (see Fig. \ref{fig:sO2_net_arch_UDA}) that was trained with an 2-norm loss function for 100 epochs on 100 2D \emph{NA} images, with a validation set of 100 images. Adam was used as the optimizer with an initial learning rate of 0.0001. A test set of 50 \emph{NA} images belonging to the same distribution as the training data were used to demonstrate the network's performance (see Fig. \ref{fig:UDA_sO2_net_evaluation} for the distribution of sO$_2$ estimates). A step by step description of the training and evaluation are given below.

\begin{enumerate}
    \item 2D images were extracted from the 60th slice of 600 3D \emph{NA} images (40x120x120x4) simulated using the light model and tissue model construction procedures outlined in \cite{bench2020toward}. These were the same 600 images used to construct the \emph{NA} training set for the CycleGAN. 
    \item The training data was normalised by dividing all 600 (40x120x4) images by the difference between their max and min values. 
    \item 100 examples were used as a training set, and 100 additional examples for a validation set. The images were cropped to have dimensions of 40x40x2 (wavelength slices corresponding to 784 nm and 820 mn wavelengths, and slices 60-100 for the second dimension).
    \item The sO$_2$ net was trained with a batch size of 5. Training was terminated when the validation loss ceased to decrease (this occurred at 100 epochs).
    \end{enumerate}
    \textbf{sO$_2$-net evaluation on adapted $A$ images}
    \begin{enumerate}
    \item  Adapted $A$ images were renormalised for use as inputs to the trained sO$_2$-net (adapted output by the CycleGAN are normalised but in the wrong way for processing with the sO$_2$-net).
    \begin{itemize}
        \item  First, the CycleGAN normalisation was inverted by multiplying the images by the difference between the max and min of all 600 available cropped \emph{NA} CycleGAN training images, and then adding the mean.
        \item Then the images were normalised for the sO$_2$ net by dividing the difference between the max and the min of all 600 uncropped (40x120x4) \emph{NA} image examples. 
    \end{itemize}
    \item The re-normalised adapted images were then input into the sO$_2$-net for evaluation.
\end{enumerate}

\label{sec:cycleGAN_sO2_train}

\newpage
\newpage
\subsection{CycleGAN experiment: additional data}
\label{sec:cycleGAN_additional_data}
\begin{figure*}[h!]
    \centering
    \includegraphics[width = .8\columnwidth]{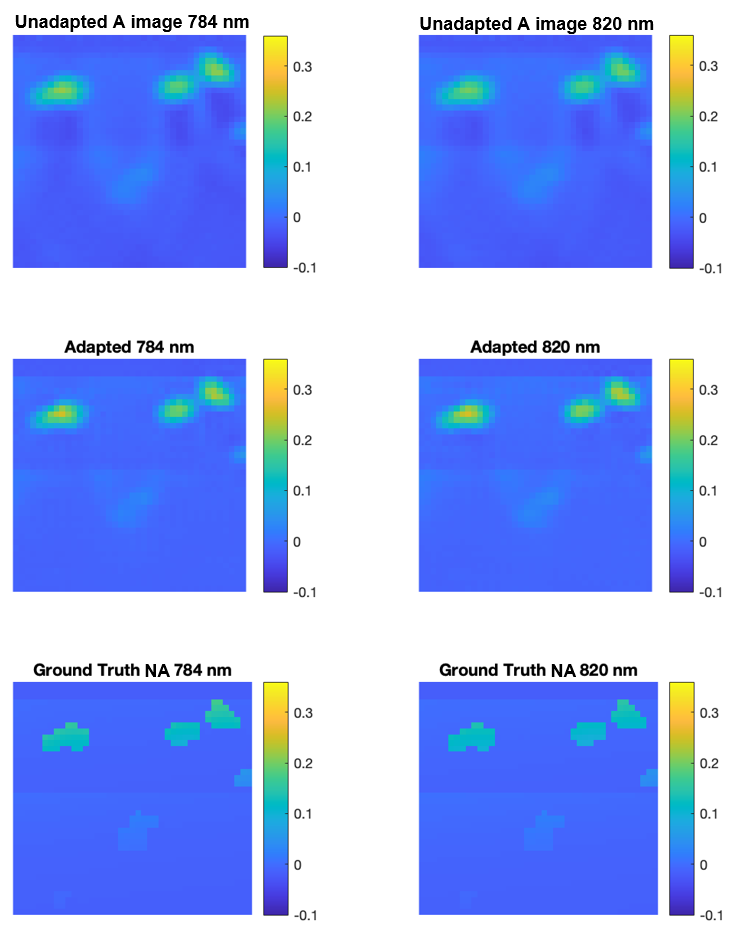}
    \caption{Example of two unadapted $A$ images (top row), their adapted counterparts appearing as though they belong to the \emph{NA} domain (middle row), and the corresponding `true' \emph{NA} versions of the images (bottom row). The CycleGAN has adapted the $A$ images to remove their wing-like artefacts and regions of low amplitude underneath the vessels, while preserving much of the original tissue's structure.}
    \label{fig:cycleGAN_adapted_ex_2}
\end{figure*}

Here, we provide some additional data from the CycleGAN domain adaptation experiment. Fig. \ref{fig:cycleGAN_adapted_ex_2} shows an additional example of an adapted set of images. 

\ifCLASSOPTIONcaptionsoff
  \newpage
\fi

\end{document}